\documentstyle[12pt,aaspp4,psfig]{article}
\def\gro04{GRO~J0422$+$32}
\def\ea{\hbox{et~al.}}
\lefthead{F. van der Hooft \ea}
\righthead{Hard X-ray variability GRO~J0422$+$32}

\begin{document}

\title{Hard X-ray variability of the black-hole candidate GRO~J0422$+$32 during its 1992 outburst} 

\author{
F.~van~der~Hooft\altaffilmark{1},
C.~Kouveliotou\altaffilmark{2,3},
J.~van~Paradijs\altaffilmark{1,4},
W.S.~Paciesas\altaffilmark{4},
W.H.G.~Lewin \altaffilmark{5},
M.~van~der~Klis\altaffilmark{1,6},
D.J.~Crary\altaffilmark{3,7},
M.H.~Finger\altaffilmark{2,3},
B.A.~Harmon \altaffilmark{3},
and S.N.~Zhang\altaffilmark{2,3}
}

\altaffiltext{1}{Astronomical Institute ``Anton Pannekoek'', University of Amsterdam
and Center for High Energy Astrophysics, Kruislaan 403, NL-1098 SJ Amsterdam, The Netherlands}
\altaffiltext{2}{Universities Space Research Association, Huntsville, AL 35806, USA} 
\altaffiltext{3}{ES-84, NASA/Marshall Space Flight Center, Huntsville, AL 35812, USA} 
\altaffiltext{4}{Department of Physics, University of Alabama in Huntsville, Huntsville, AL 35899, USA} 
\altaffiltext{5}{Massachusetts Institute of Technology, 37-627 Cambridge, MA 02139, USA}
\altaffiltext{6}{Department of Astronomy, University of California, Berkeley, Berkeley, CA 94720, USA}
\altaffiltext{7}{NAS/NRC Research Associate, NASA Code ES-84, Marshall Space Flight Center, Huntsville, AL 35812, USA}

%
%================== Abstract ===============================
%
\begin{abstract} 

We have studied the hard X-ray variability of the soft X-ray transient \gro04\ with BATSE in the 
20--100 keV energy band. 
Our analysis covers 180 days following the first X-ray detection of the source on 1992 August 5, 
fully covering its primary and secondary X-ray outburst. 
We computed power density spectra (PDSs) in the 20--50, 50--100, and 20--100 keV energy bands, in 
the frequency interval 0.002--0.488 Hz. 
The PDSs of \gro04\ are approximately flat up to a break frequency, and decay as a power law above, 
with index $\sim$\,1. 
During the first 70 days of the X-ray outburst, the PDSs of \gro04\ show a significant QPO peak near 
$\sim$\,0.2 Hz, superposed on the power-law tail. 

The break frequency of the PDSs obtained during the primary X-ray outburst of \gro04\ occurs at 
0.041$\pm$0.006 Hz; during the secondary outburst the break is at 0.081$\pm$0.015 Hz. 
The power density at the break ranged between 44 and 89\% Hz$^{-1/2}$ (20--100 keV). 
The canonical anticorrelation between the break frequency and the power density at the break, observed in Cyg~X-1 
and other BHCs in the low state, is not observed in the PDSs of \gro04. 

We compare our results with those of similar variability studies of Cyg~X-1. 
The relation between the spectral slope and the amplitude of the X-ray variations of \gro04\ is similar 
to that of Cyg~X-1; however, the relation between the hard X-ray flux and the amplitude of its variation 
is opposite to what has been found in Cyg~X-1. 
Phase lags between the X-ray flux variations of \gro04\ at high and low photon energies, could only be 
derived during the first 30 days of its outburst. 
During this period, the variations in the 50--100 keV, lag those in the 20--50 keV energy band 
by an approximately constant phase difference of 0.039(3) rad in the frequency interval 0.02--0.20 Hz. 
The time lags of \gro04\ during the first 30 days of the outburst, decrease with frequency as a power law, 
with index 0.9 for $\nu$\,$>$\,0.01 Hz.
\end{abstract} 


%
%================== Text and Acknowledgements ===============================
%

\section{Introduction} 

Soft X-ray transients (SXTs) are low-mass X-ray binaries consisting of a
neutron star or black-hole primary that undergoes unstable accretion from a
late-type companion. 
Disk instabilities (van Paradijs 1996; King, Kolb \& Burderi 1996) 
cause brief but violent outbursts, typically lasting weeks to months, 
during which the X-ray luminosity abruptly increases by several orders of magnitude, 
to $\sim$\,$10^{37}$--5$\times 10^{38}$ erg sec$^{-1}$, separated by quiescent intervals 
lasting from years to many decades. 

\begin{figure*} 
\centerline{\psfig{figure=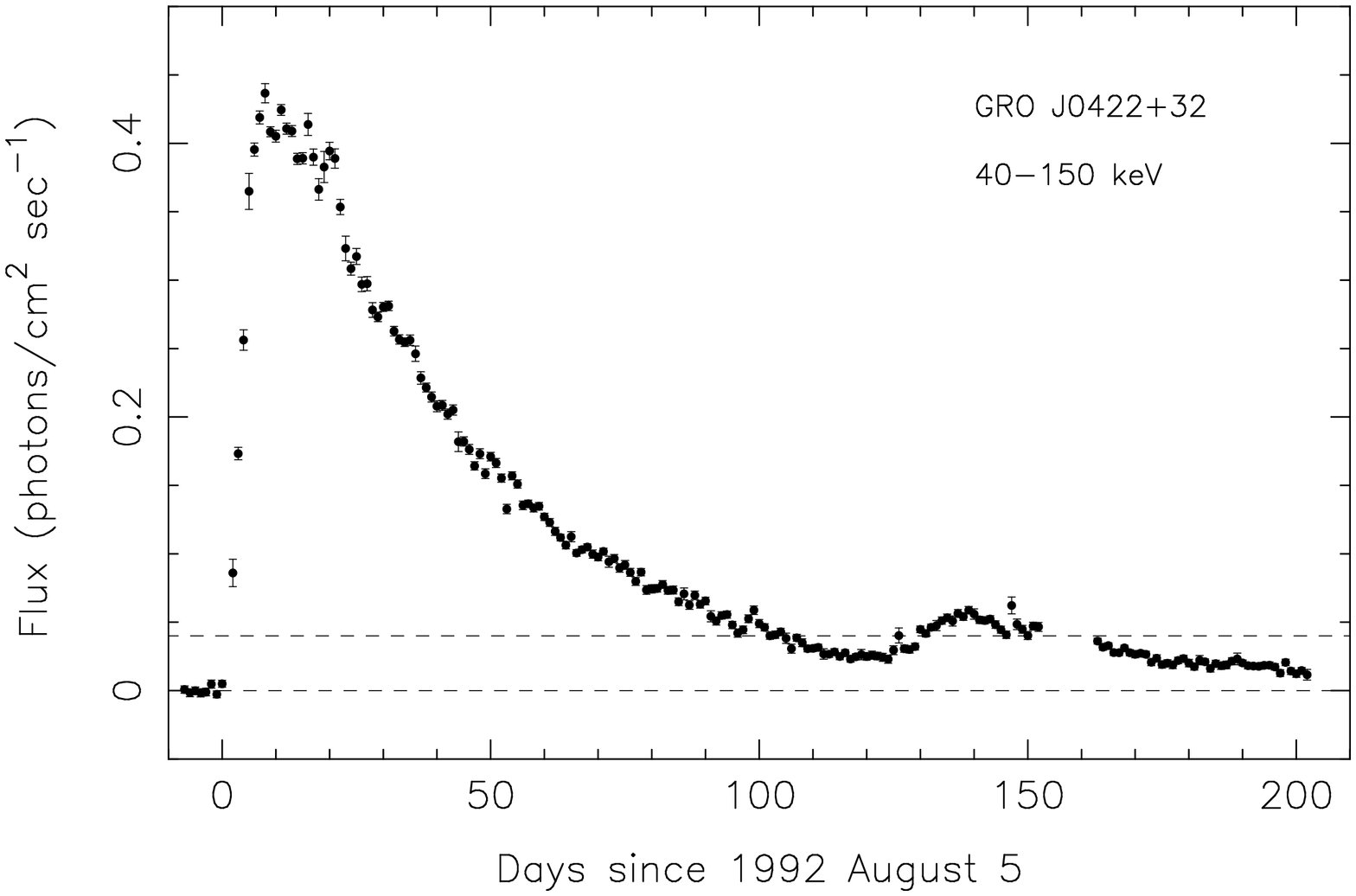,width=12cm,bbllx=159pt,bblly=85pt,bburx=547pt,bbury=687pt,angle=-90}}
\caption[]{
Flux history of \gro04\ measured by BATSE in the 40--150 keV energy band. The source was detected 
first on 1992 August 5 and reached a maximum flux of $\sim$\,3 Crab within 3 days. Hereafter the 
X-ray intensity decreased exponentially, and reached a secondary (local) maximum in X-ray intensity 
at 139 days after its initial detection. The source became undetectable by BATSE after 
$\sim$\,200 days. Dashed lines indicate the levels 0.00 and 0.04 photons/cm$^{\rm 2}$ sec$^{-1}$.
} 
\label{0422_fig1} 
\end{figure*}

The SXT \gro04\ (Nova Persei 1992) was detected 
with the Burst And Transient Source Experiment (BATSE) on board the Compton 
Gamma Ray Observatory on 1992 August 5 (Paciesas \ea\ 1992). 
Initially, the hard X-ray flux of the source increased rapidly, reaching a maximum of 
$\sim$\,3 Crab (20--300 keV) within three days after first detection (Paciesas \ea\ 1992), 
at which level it remained for the following three days (Harmon \ea\ 1992). 
Subsequently, the hard X-ray intensity (40--150 keV) of \gro04\ decreased 
exponentially with a decay time of 43.6 days (Vikhlinin \ea\ 1995). 
About 135 days after the primary X-ray maximum, the X-ray flux of \gro04\ 
reached a secondary maximum, after which the flux continued to decrease with 
approximately the same decay time as before. 
The source was detected above the BATSE 3$\sigma$ one-day detection treshold of 0.1 Crab 
(20--300 keV) for $\sim$\,200 days following the start of the X-ray outburst. 
The daily averaged 40--150 keV flux history of \gro04\ during this period is displayed in 
Figure~\ref{0422_fig1}.

The X-ray spectrum of \gro04\ was hard and could be well described by a cut-off power law with photon 
index $\sim$\,1.5 and break energy $\sim$\,60 keV, detected up to 600 keV with OSSE 
(Grove, Kroeger \& Strickman 1997). 
In observations of \gro04\ during X-ray maximum with COMPTEL, the source was detected 
up to 1--2 MeV (van Dijk \ea\ 1995). 
Timing analysis of the hard X-ray data revealed quasi-periodic oscillations (QPOs) 
centered at 0.03 and 0.2 Hz (20--300 keV BATSE data, Kouveliotou \ea\ 1992, 1993)  
and 0.3 Hz (40--150 keV SIGMA data, Vikhlinin \ea\ 1995). 
These QPOs detected in hard X-rays were confirmed by ROSAT observations in the 
0.1--2.4 keV energy band (Pietsch \ea\ 1993). 
The power density spectra (PDSs) of \gro04\ were flat for frequencies, {\it f}, below the 
first QPO peak at 0.03 Hz, and fall as 1/{\it f} between both QPOs (Kouveliotou \ea\ 1992).  
The total fractional rms variations of the 40--150 keV flux in the 10$^{-3}$--10$^{-1}$ Hz 
frequency interval, ranged between $\sim$\,15\% and $\sim$\,25\% (Vikhlinin \ea\ 1995). 
The observed hard X-ray spectrum and rapid X-ray variability resemble the properties of 
dynamically proven black-hole candidates (BHCs) (Sunyaev \ea\ 1991; Tanaka \& Lewin 1995; 
see however, van Paradijs \& van der Klis 1994), 
which led to the suggestion that \gro04\ is also a BHC (Roques \ea\ 1994).

Soon after its first X-ray detection, the optical counterpart of \gro04\ 
was independently identified by Castro-Tirado \ea\ (1992, 1993)  
and Wagner \ea\ (1992)  
at a peak magnitude of {\sl V}\,$=$\,13.2. This object was absent on the POSS down to a limiting 
magnitude of {\sl R}\,$\simeq$\,20 (Castro-Tirado \ea\ 1993).  
During the first 210 days of the X-ray outburst the optical light curve declined exponentially 
with an {\sl e}--folding time of 170 days (Shrader \ea\ 1994). 
Then the optical brightness dropped quickly, followed by two mini-outbursts 
(with an amplitude of $\sim$\,4 mag each) before it finally reached 
quiescence at {\sl V}\,$=$\,22.35, $\sim$\,800 days after first detection of the 
X-ray source (Garcia \ea\ 1996). 
Therefore, the outburst amplitude of \gro04\ was about 9 mag in {\sl V}, the largest 
observed in any SXT to date (van Paradijs \& McClintock 1995).  
The optical light curve of \gro04, during and after the X-ray outburst, bears many characteristics 
similar to those of the so-called ``tremendous outburst amplitude dwarf novae'' (TOADs). 
Kuulkers, Howell \& van Paradijs (1996) 
proposed that these similarities reflect the small mass ratios and very low mass transfer rates 
in SXTs and TOADs.

During the decay to quiescence, optical brightness modulations were reported 
at periods of $\sim$\,2.1, 5.1, 10.2, and 16.2 hrs, respectively (Harlaftis \ea\ 1994; 
Callanan \ea\ 1995; Kato, Mineshige \& Hirata 1995; Chevalier \& Ilovaisky 1995; 
Martin \ea\ 1995).  
Spectroscopic observations by Filippenko, Matheson \& Ho (1995) 
showed that the orbital period is 5.08$\pm$0.01 hrs, and determined the mass function at 
{\it f\,(M)}\,$=$\,1.21$\pm$0.06 M$_{\odot}$. This value of the mass function is confirmed by observations 
performed by Orosz \& Bailyn (1995), {\it f\,(M)}\,$=$\,0.40--1.40 M$_{\odot}$, and Casares \ea\ (1995), 
{\it f\,(M)}\,$=$\,0.85$\pm$0.30 M$_{\odot}$. 
The orbital inclination, $i$, of \gro04\ was estimated 
from an interpretation of the double waved light curve as an ellipsoidal variation. 
From the $\Delta${\sl I}\,$\sim$\,0.03 mag semi-amplitude modulation Callanan \ea\ (1996) 
derived $i$\,$\leq$\,45$^\circ$. 
This implies a mass of $\geq$\,3.4 M$_{\odot}$ for the compact star, i.e. slightly 
above the (theoretical) maximum mass of a neutron star. 
Independently, Beekman \ea\ (1997) constrain the inclination of \gro04\ from infrared and optical 
photometry to be between 10$^\circ$ and 31$^\circ$. Consequently, their lower limit to the mass 
of the compact star becomes $\ga$\,9 M$_{\odot}$. 
Therefore, the compact star in \gro04\ is most probably a black hole. 

Here we report on a temporal analysis of 20--100 keV BATSE data of \gro04\ collected 
during 180 days following its first detection. We derive the X-ray light curve of the source and 
discuss its hard X-ray spectrum. We compute daily averaged power density spectra in the frequency 
interval 0.002--0.488 Hz and show their evolution in time. From the Fourier amplitudes we compute 
cross spectra and derive the phase and time lags between the hard and soft X-ray variations of \gro04. 
We compare our results with those of similar analyses of Cyg~X-1.

\section{X-ray light curve and X-ray spectrum}
\label{lc_spectrum}

We have determined the flux of \gro04\ by applying the Earth occultation technique (Harmon \ea\ 1993) 
to data obtained with the BATSE large-area detectors (count rates with a 2.048 sec time resolution in 16 spectral 
channels).
The average number of occultation steps of \gro04\ was 13 per day, ranging from 2 to 24. 
The systematic error due to the variable orientation of {\sl CGRO}, is typically $\sim$\,5--10\%. 
The daily averaged flux history of \gro04\ in the 40--150 keV energy band is displayed in Fig.~\ref{0422_fig1}. 
The X-ray light curve rises from first detection to a maximum of $\sim$\,3 Crab (20--300 keV) in 
three days, followed by an exponential decrease with a decay time of 43.6 days (Vikhlinin \ea\ 1995). 
During the decay of the X-ray light curve, a secondary maximum is observed at about 135 days after the main 
X-ray maximum. 
After the secondary maximum the light curve decays at approximately the same rate as before. 
The whole 1992 X-ray outburst of \gro04 lasted for $\sim$\,200 days. 

\begin{figure*} 
\centerline{\psfig{figure=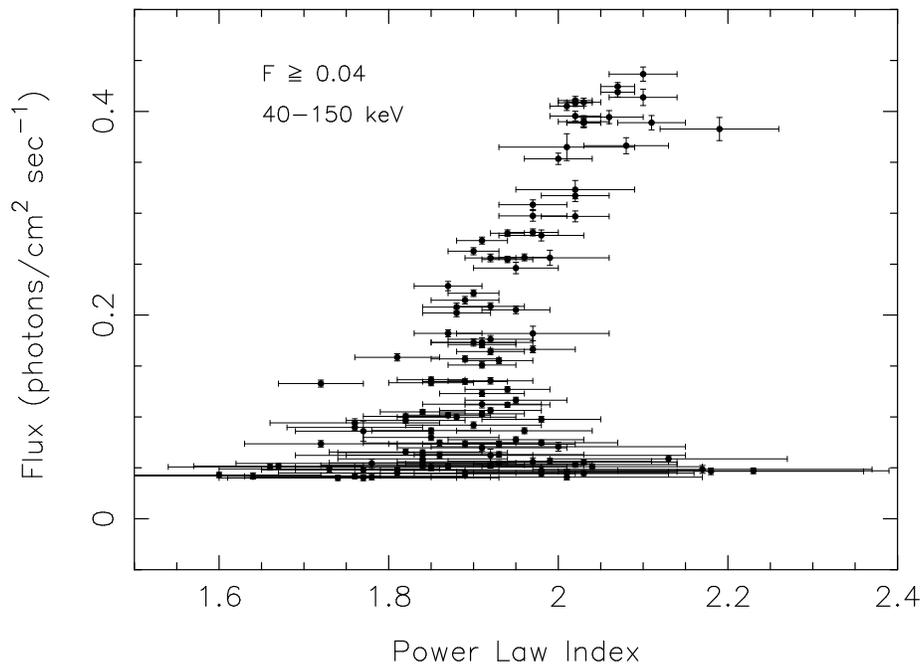,width=12cm,bbllx=153pt,bblly=85pt,bburx=548pt,bbury=620pt,angle=-90}}
\caption[]{
Distribution of the daily averaged flux measurements versus photon power-law index of \gro04\ in the 
40--150 keV energy band. The sample consists of daily averaged data and is flux-limited 
({\it F}$\geq$0.04 photons/cm$^{\rm 2}$ sec$^{-1}$). The distribution consists of 103 data points taken 
during the primary outburst, and 23 days of the secondary outburst.
} 
\label{0422_fig2} 
\end{figure*}

The 40--150 keV X-ray spectrum of \gro04\ measured by BATSE was approximated by a single power law. 
A single power law does not provide a good fit to the X-ray spectrum, but is useful as a coarse indicator 
of the spectral hardness. 
The power-law index of the spectrum was $\sim$\,2.1 at the peak of the 
primary outburst; the spectrum became harder during its decay (power-law index of $\sim$\,1.9). The 
distribution of one-day averaged flux measurements versus the photon power-law indices of \gro04\ in the 
40--150 keV energy band is shown in Figure~\ref{0422_fig2}. The distribution is flux-limited 
($F\geq0.04$ photons/cm$^{\rm 2}$ sec$^{-1}$) and contains a total of 126 data points, 23 of which 
corresponding to the secondary outburst. For flux values above 0.20 photons/cm$^{\rm 2}$ sec$^{-1}$, 
flux and power-law index appear to be correlated positively. 
For $F\leq0.20$ photons/cm$^{\rm 2}$ sec$^{-1}$, the photon  
power-law index is determined less accurately, and remains constant at 1.89$\pm$0.02 (average of 84 data points, 
weighted by the individual errors). The weighted average of the photon power-law index obtained during 
the primary outburst only (1.89$\pm$0.02, 61 data points), is consistent with that of 
the secondary outburst (1.92$\pm$0.04, 23 data points).

\section{Time series analysis}

We have used count rate data from the large-area detectors (four broad energy channels) 
with a time resolution of $\Delta${\it T}\,$=$\,1.024 sec and applied an empirical model 
(Rubin \ea\ 1996) 
to subtract the signal due to the X-ray/gamma-ray background. 
This model describes the background by a harmonic expansion in 
orbital phase (with parameters determined from the observed background 
variations), and includes the risings and settings of the brightest X-ray 
sources in the sky. It uses eight orbital harmonic terms, and its parameters 
were updated every three hours.

For our analysis we considered data segments of 524.288 sec (512 time bins of 
1.024 sec each) on which we performed Fast Fourier Transforms (FFTs) covering 
the frequency interval 0.002--0.488 Hz. The average number of uninterrupted 512 bin 
segments available while the source was not occulted by the Earth was 32 per day. 
For each of the detectors which had the source within 60$^\circ$ of the normal, 
we calculated the FFTs of the lowest two energy channels (20--50, 50--100 keV) of each 
data segment separately. 
These FFTs were coherently summed (weighted by the ratio of the source to the 
total count rates) and converted to PDSs. 
The source count rates were obtained from the occultation analysis. 
The PDSs were normalized such that the power density is given in units of 
(rms/mean)$^{2}$ Hz$^{-1}$ (see, e.g., van der Klis 1995a) 
and finally averaged over an entire day.

\subsection{Power density spectra (PDSs)}

\begin{figure*} 
\centerline{\psfig{figure=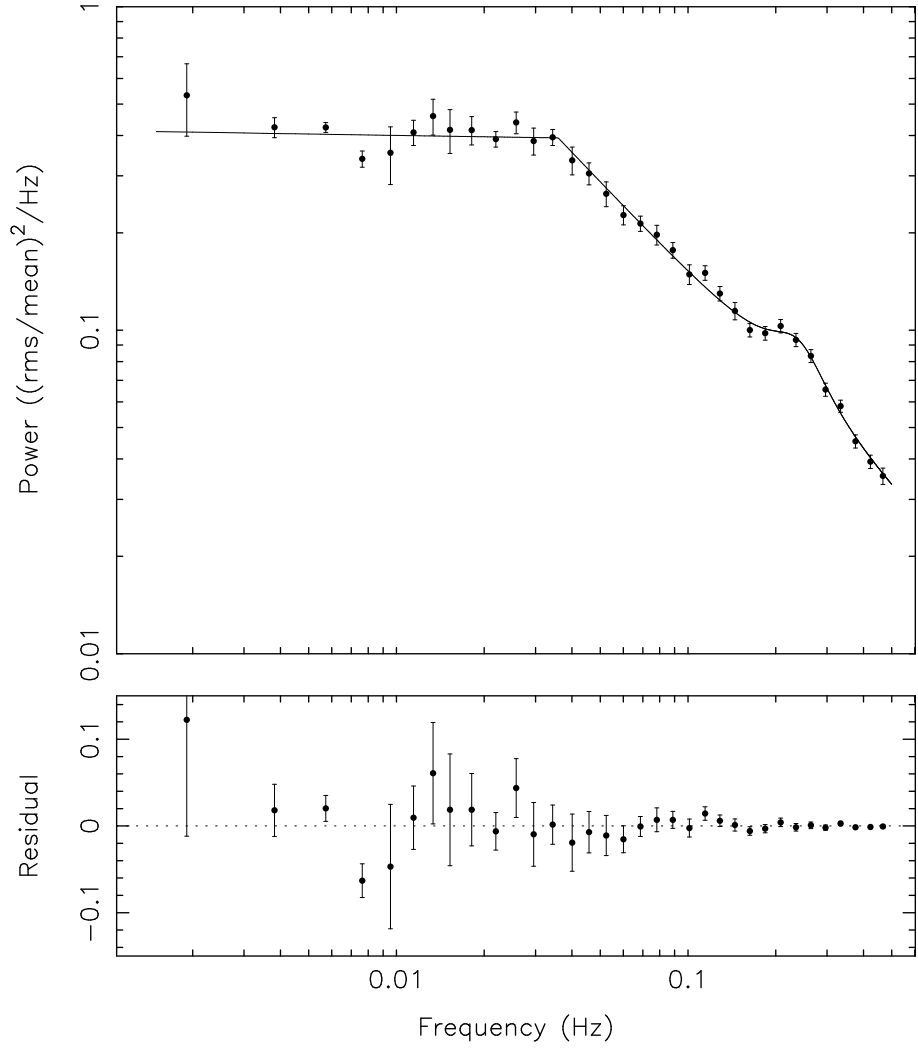,width=12cm,bbllx=39pt,bblly=222pt,bburx=520pt,bbury=778pt}}
\caption[]{
The five-day averaged power spectrum (20--100 keV) of \gro04\ for days 33--37 and the best-fit model 
(broken power law plus Lorentzian profile) {\sl (top panel)}, and the residuals {\sl (bottom panel)}. 
The frequency scale has been logarithmically rebinned into 34 bins.
}
\label{0422_fig4} 
\end{figure*}

The power spectra of \gro04\ are approximately flat up to a break frequency, 
and decay as a power law above (see Figure~\ref{0422_fig4}). 
During the first $\sim$\,70 days of the outburst, 
the PDSs show a significant peak, indicative of QPOs in the time series, superposed on 
the power-law tail. 
For all PDSs obtained during this stage of the X-ray outburst, the centroid frequency 
of the QPO peak is above the break frequency. 
Beyond day $\sim$\,70, the QPOs are no longer detected significantly in the PDSs; the power 
spectra obtained during the remaing part of the outburst are well described by a broken power law only. 

\begin{figure*} 
\centerline{\psfig{figure=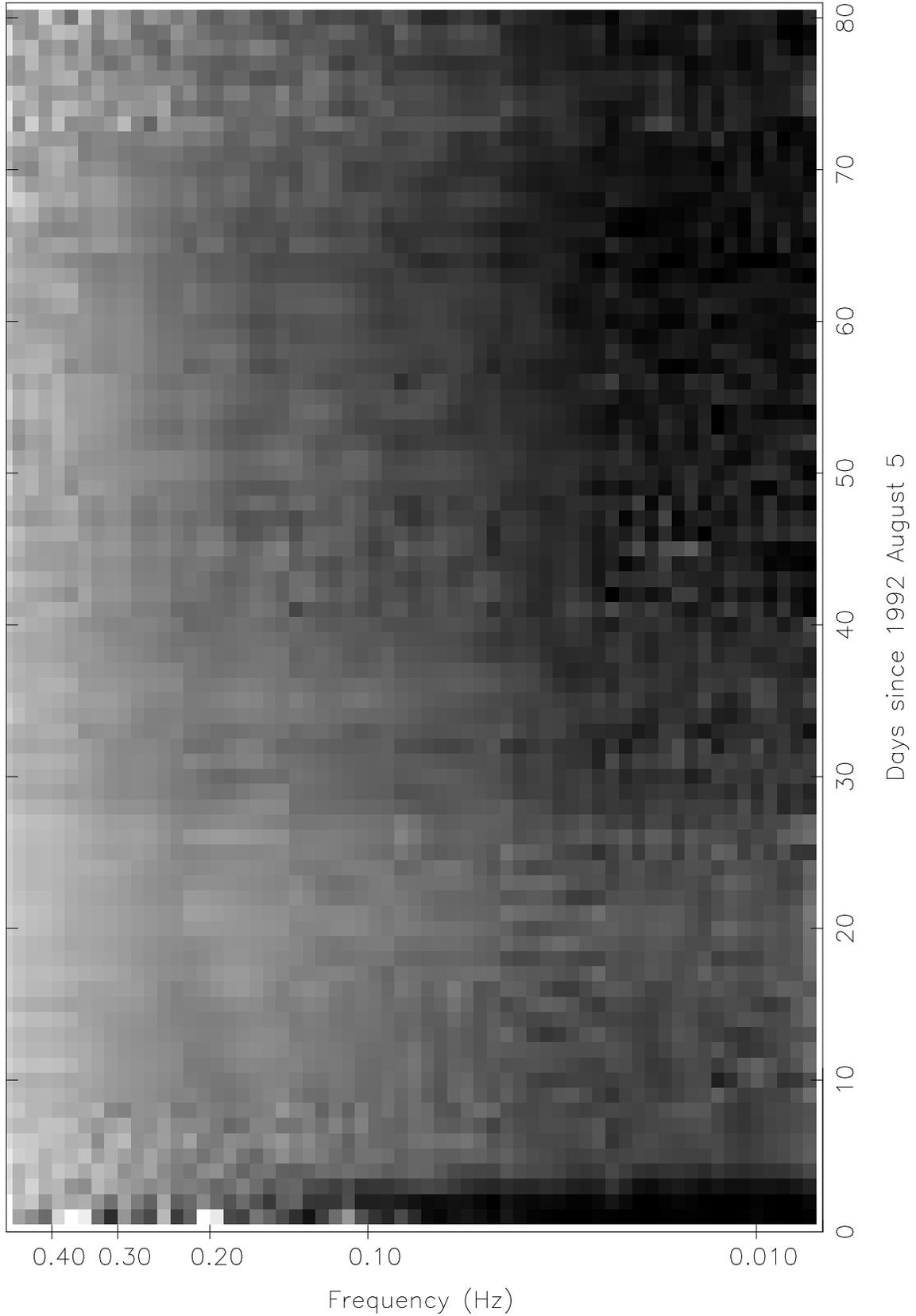,height=22cm,bbllx=25pt,bblly=59pt,bburx=529pt,bbury=784pt}} 
\caption[]{
Dynamical spectrum of the set of daily averaged, (rms/mean)$^{2}$ Hz$^{-1}$ 
normalized PDSs (20--100 keV) covering the first 80 days of the outburst of \gro04. 
The frequency scale has been logarithmically rebinned; dark colors indicate a high power level. 
}
\label{0422_fig3} 
\end{figure*}

\begin{figure*} 
\centerline{\psfig{figure=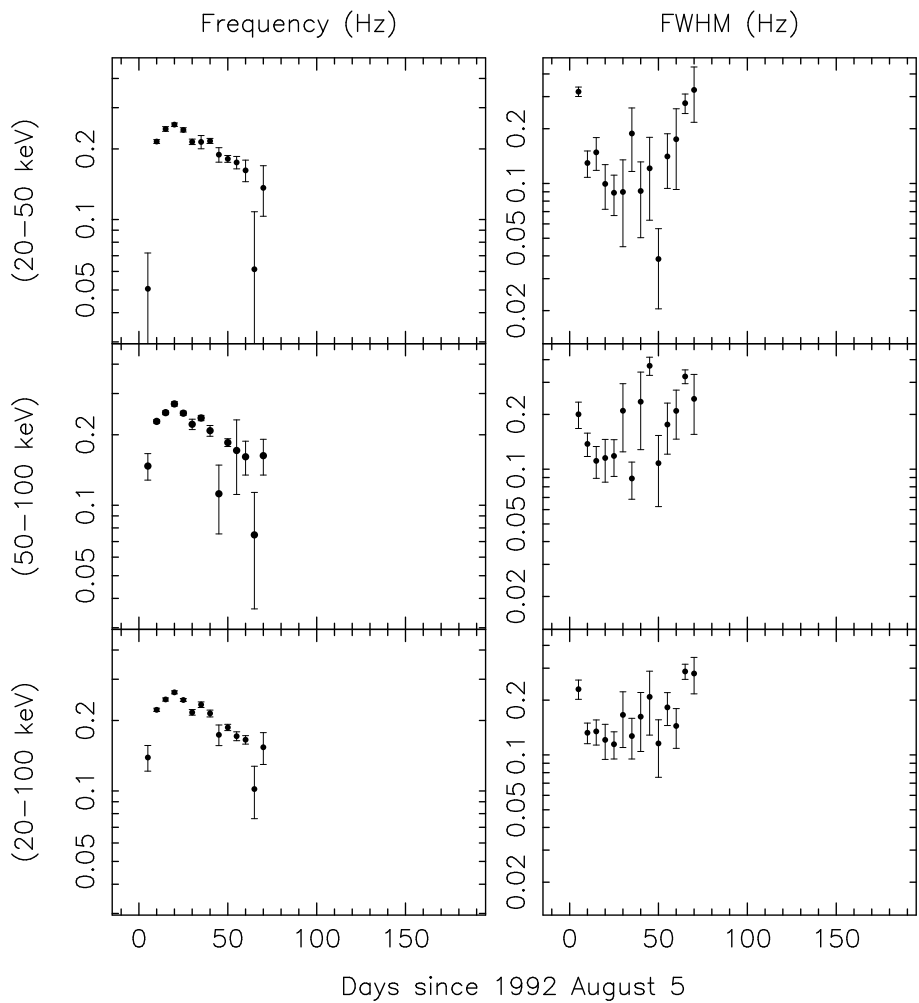,width=12cm,bbllx=40pt,bblly=66pt,bburx=565pt,bbury=540pt,angle=-90}} 
\caption[]{
Parameters of Lorentzian profile of the model fit to the five-day averaged PDSs of \gro04\ 
in the 20--50 {\sl (top row)}, 50--100 {\sl (middle row)}, and 20--100 keV {\sl (bottom row)} 
energy bands. The history of the centroid frequency {\sl (left column)}, and full width at half 
maximum {\sl (right column)} of the Lorentzian are displayed logarithmically. 
The Lorentzian component of the fit was included during the first 70 days of the outburst of \gro04\ only.
} 
\label{0422_fig6}
\end{figure*}

The evolution of the daily averaged PDSs (20--100 keV) of \gro04\ during the first 80 days of the 
outburst is illustrated in a dynamical spectrum (Figure~\ref{0422_fig3}). 
In this representation the frequency scale is 
logarithmically rebinned to 61 bins; the darker the color, the higher is the power level. 
In Fig.~\ref{0422_fig3}, a dark shaded band, indicative of QPOs in the time series, is 
seen at centroid frequencies between 0.13 and 0.26 Hz until approximately day 70 of the outburst. 
The centroid frequency of the QPO increases slowly until day 20, and gradually decreases 
afterwards (see also Figure~\ref{0422_fig6}). 
During days 50--70 of the outburst, there is some evidence for the 
presence of a second, very weak peak in the power spectra. This second feature is 
much weaker than the main QPOs peak, and is found at frequencies between the 
break of the power law and the centroid frequency of the main QPO. 
From about 70 days after the start of the outburst onward, neither the main QPOs peak nor 
the second feature are present significantly in the PDSs of \gro04.

The break in the power spectra of \gro04\ is clearly visible at low frequencies in the dynamical 
spectrum of Fig.~\ref{0422_fig3}. 
During the period displayed in this dynamical spectrum, the break frequency remains constant, 
and occurs at 0.037$\pm$0.004 Hz.
As the outburst of \gro04\ proceeds, the power density increases over the total frequency 
interval of the PDSs. 
This effect is most prominently seen at low frequencies in the dynamical spectrum, 
indicated by the darker shaded area starting at day $\sim$\,30, but is present in the 
total frequency interval of the power spectra. 
From Fig.~\ref{0422_fig3} it also follows that the power density of the PDSs obtained during the first 
three days of the X-ray outburst of \gro04\ is enhanced with respect to the remaining data, 
indicating increased rapid variability at the very start of the outburst compared to later 
stages of the outburst.

\subsection{Fits to the PDSs}

We made fits to the PDSs in the 20--50, 50--100 and 20--100 keV energy bands, using a 
combination of a Lorentzian profile and a broken power law. To improve statistics, 
we made fits to the average PDSs of five consecutive days during the first 110 days of the outburst 
(obtaining 22 five-day averaged PDSs), and averages of ten consecutive days afterward 
(obtaining 7 ten-day averaged PDSs). A typical five-day averaged power spectrum in the 
20--100 keV energy band, together with the model fit, is shown in the top panel of 
Fig.~\ref{0422_fig4}. 
The model requires seven parameters (three for the Lorentzian, four for the broken power 
law), which left 27 degrees of freedom as the PDSs were logarithmically rebinned into 34 
frequency bins. The Lorentzian profile was included during the first 70 days of the outburst 
only (i.e. 14 five-day averaged PDSs). Inclusion of a Lorentzian profile during later stages of 
the outburst did not significantly improve the quality of the fit. We routinely obtained 
reduced $\chi^{2}$ values between 0.6 and 2.9 for the fits to the PDSs. 
As the outburst of \gro04\ continued, the received flux decreased (see, e.g., Fig.~\ref{0422_fig1}) 
for which reason the power spectra became ill-constrained at the local flux minimum near day 
$\sim$\,120 of the outburst. During the secondary maximum the quality of the PDSs improved slightly. 

\begin{figure*} 
\centerline{\psfig{figure=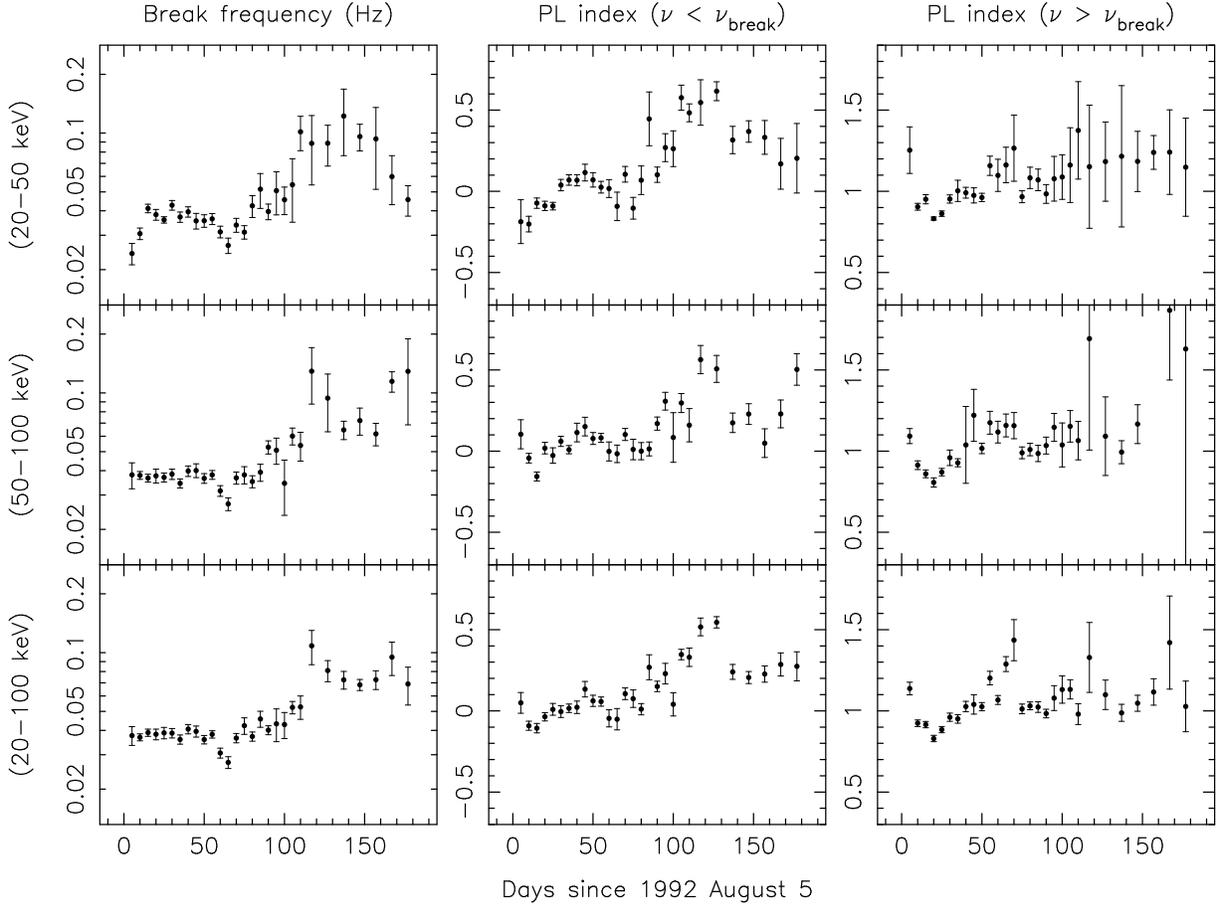,width=16cm,bbllx=40pt,bblly=66pt,bburx=565pt,bbury=766pt,angle=-90}} 
\caption[]{
Parameters of broken power-law component of the model fit to the five and ten-day averaged PDSs of \gro04\ 
in the 20--50 {\sl (top row)}, 50--100 {\sl (middle row)}, and 20--100 keV {\sl (bottom row)} 
energy bands. The history of the break frequency $\nu_{\rm break}$ {\sl (left column)}, the power-law 
index for $\nu < \nu_{\rm break}$ {\sl (middle column)}, and the power-law index for $\nu > \nu_{\rm break}$ 
{\sl (right column)} of the broken power law are displayed. The Lorentzian component of the fit was included 
during the first 70 days of the outburst of \gro04\ only (i.e. first 14 data points).
} 
\label{0422_fig5}
\end{figure*}

The histories of the break frequency $\nu_{\rm break}$, and indices of the broken power-law component 
for $\nu < \nu_{\rm break}$, and $\nu > \nu_{\rm break}$ (29 data points in each energy band), 
are presented in Figure~\ref{0422_fig5}. 
The evolution of the centroid frequency and full width at half maximum (FWHM) of the Lorentzian profile 
(14 data points per energy band only, as the Lorentzian profile was included only during the first 
70 days of the X-ray outburst) are shown in Fig.~\ref{0422_fig6}. 
From these figures it can be seen that the centroid frequency of the Lorentzian profile increased during 
the first 3 weeks of the outburst of \gro04 from an initial value of $\sim$\,0.15 Hz to a maximum of $\sim$\,0.26 Hz, 
but from that moment on monotonically decreased to $\sim$\,0.1 Hz 70 days after the onset of the outburst. 
The FWHM of the Lorentzian is determined best in the PDSs in the 20--100 keV energy band. 
The FWHM of the QPO peak in this energy band does not change significantly during the outburst, although there may 
be a slight slight tendency to increase as the significance of the QPOs peak in the power spectra diminishes towards 
the end of the 70 days period. 
During the first 100 days of the outburst, the average break frequency is 0.041\,$\pm$\,0.006 Hz (20--100 keV), 
followed by an increase (within less than 2 weeks) to an average value of 0.081\,$\pm$\,0.015 Hz during the last 
70 days of the outburst. 
A local minimum in the history of the break frequencies is observed near day 60 of the outburst. 
The low-frequency flat part of the PDSs below the break frequency appears to steepen during this period. 
Note that the increase of the break frequency does not coincide with the omission of the Lorentzian profile from the 
fitting function.

\subsection{Break frequency}

\begin{figure*} 
\centerline{\psfig{figure=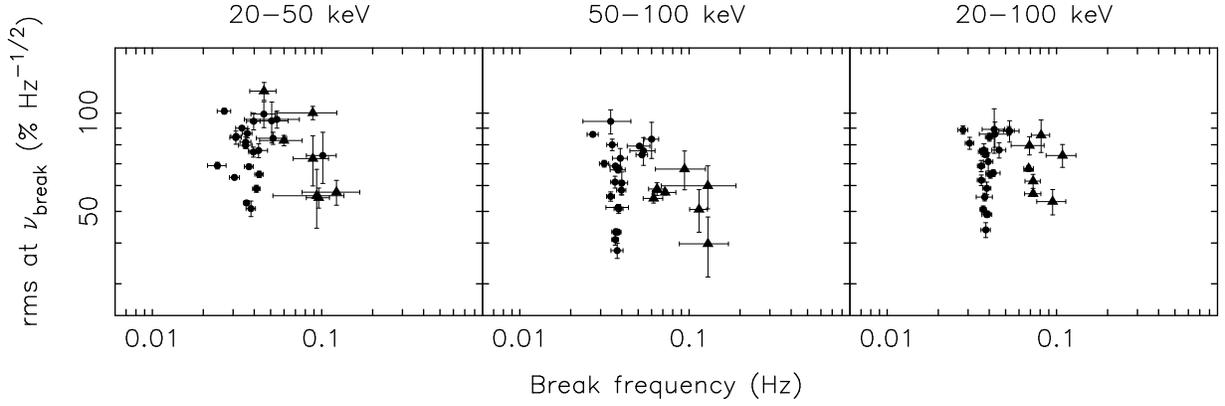,width=16cm,bbllx=47pt,bblly=80pt,bburx=263pt,bbury=722pt,angle=-90}} 
\caption[]{
Relation between break frequency and power density at the break in the PDSs of \gro04\ in the 20--50 
{\sl (left panel)}, 50--100 {\sl (middle panel)}, and 20--100 keV {\sl (right panel)} energy bands. 
The dots correspond to data derived from the 22 five-day averaged PDSs of the primary outburst of 
\gro04, while the triangles correspond 7 ten-day averaged PDSs of the following 70 days, which cover 
the secondary outburst of the source.
} 
\label{0422_fig7}
\end{figure*}

The relation between the break frequency and the power density at the break in the PDSs of \gro04, 
is displayed in Figure~\ref{0422_fig7} for the 20--50, 50--100, and 20--100 keV energy bands.
From the 20--100 keV data it follows that the power density at the break covers a broad range of values 
(44--89\% Hz$^{-1/2}$), while the break frequency appears to cluster at 0.041$\pm$0.006 and 0.081$\pm$0.015 Hz. 
Data obtained from the secondary outburst of \gro04\ (7 ten-day averaged PDSs) are indicated in 
Fig.~\ref{0422_fig7} by triangles, while those of the primary outburst (22 five-day averaged PDSs) 
are denoted by dots. 
The 20--100 keV data in this figure (right panel), clearly show that the break frequencies 
near $\sim$\,0.08 Hz all correspond to PDSs obtained during the secondary outburst of \gro04. 
Those of the primary outburst, all cluster at a frequency near $\sim$\,0.04 Hz. 
The distribution of the break frequency and the power density at the 
break determined in the 20--50 keV and the 50--100 keV data shows more scatter, but follows the 
same trend: the low break frequency corresponds to the primary outburst of \gro04, while the high break 
frequency occurs during the secondary outburst.

\begin{figure*} 
\centerline{\psfig{figure=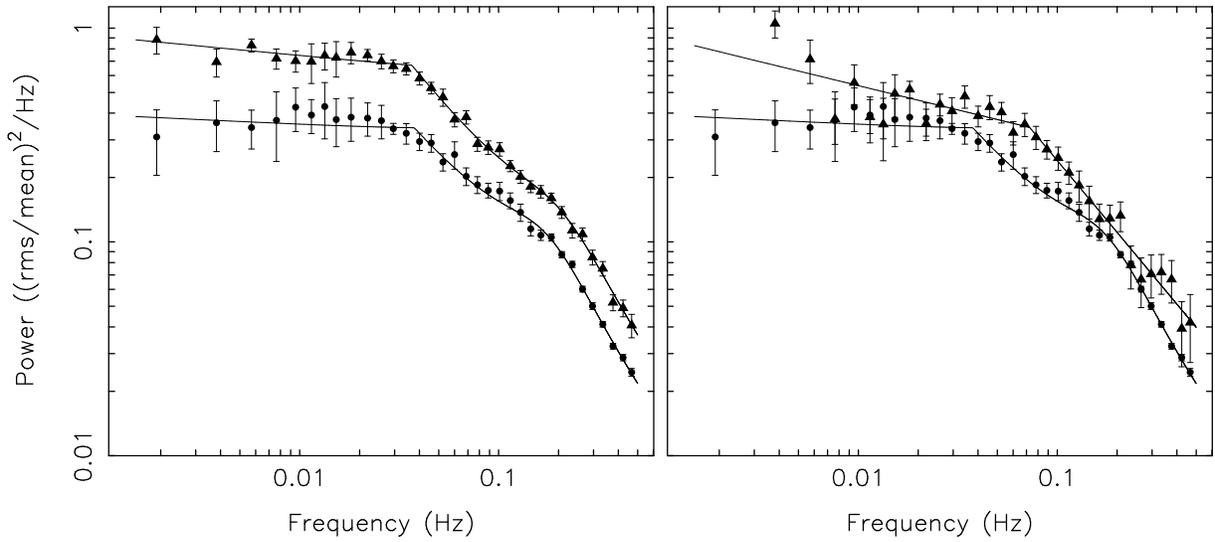,width=16cm,bbllx=239pt,bblly=91pt,bburx=536pt,bbury=757pt,angle=-90}} 
\caption[]{
The five-day averaged power PDSs (20--100 keV) of \gro04\ and the best-fit model (broken power law plus 
Lorentzian profile) for days 3--7 (dots) and 68--72 (triangles) {\sl (left panel)}, and for days 3--7 (dots) and 
153--162 (dots) (ten-day averaged PDS, fitting function consisting of a power-law component only) {\sl (right panel)}. 
The frequency scale has been logarithmically rebinned into 34 bins.
} 
\label{0422_fig8}
\end{figure*}

This effect is illustrated again in Figure~\ref{0422_fig8}, in which we show two averaged power spectra 
of \gro04\ in the 20--100 keV energy band, obtained during different stages of its outburst, as well as 
their respective model fits. 
The left panel contains two PDSs with equal break frequency, but different power density at the break, i.e., 
both PDSs were obtained during the primary outburst of \gro04. The PDS of day 3--7 (indicated by dots) 
has a break frequency and power density at the break of 
($\nu_{\rm break}$, rms)\,$=$\,(0.038$\pm$0.004 Hz, 55$\pm$2\% Hz$^{-1/2}$); triangles indicate the PDS corresponding 
to day 68--72 with (0.037$\pm$0.002 Hz, 76$\pm$2\% Hz$^{-1/2}$). 
The right panel displays two PDSs with equal power density at the break, but different break frequency, 
selected from both the primary and secondary outburst.
Dots indicate the PDS obtained during day 3--7 as in the left panel, while the triangles correspond 
to the PDS obtained during day 153--162. The break frequency and power density at the break in this 
power spectrum were (0.073$\pm$0.008 Hz, 57$\pm$1\% Hz$^{-1/2}$).

\subsection{Fractional rms amplitudes}

\begin{figure*} 
\centerline{\psfig{figure=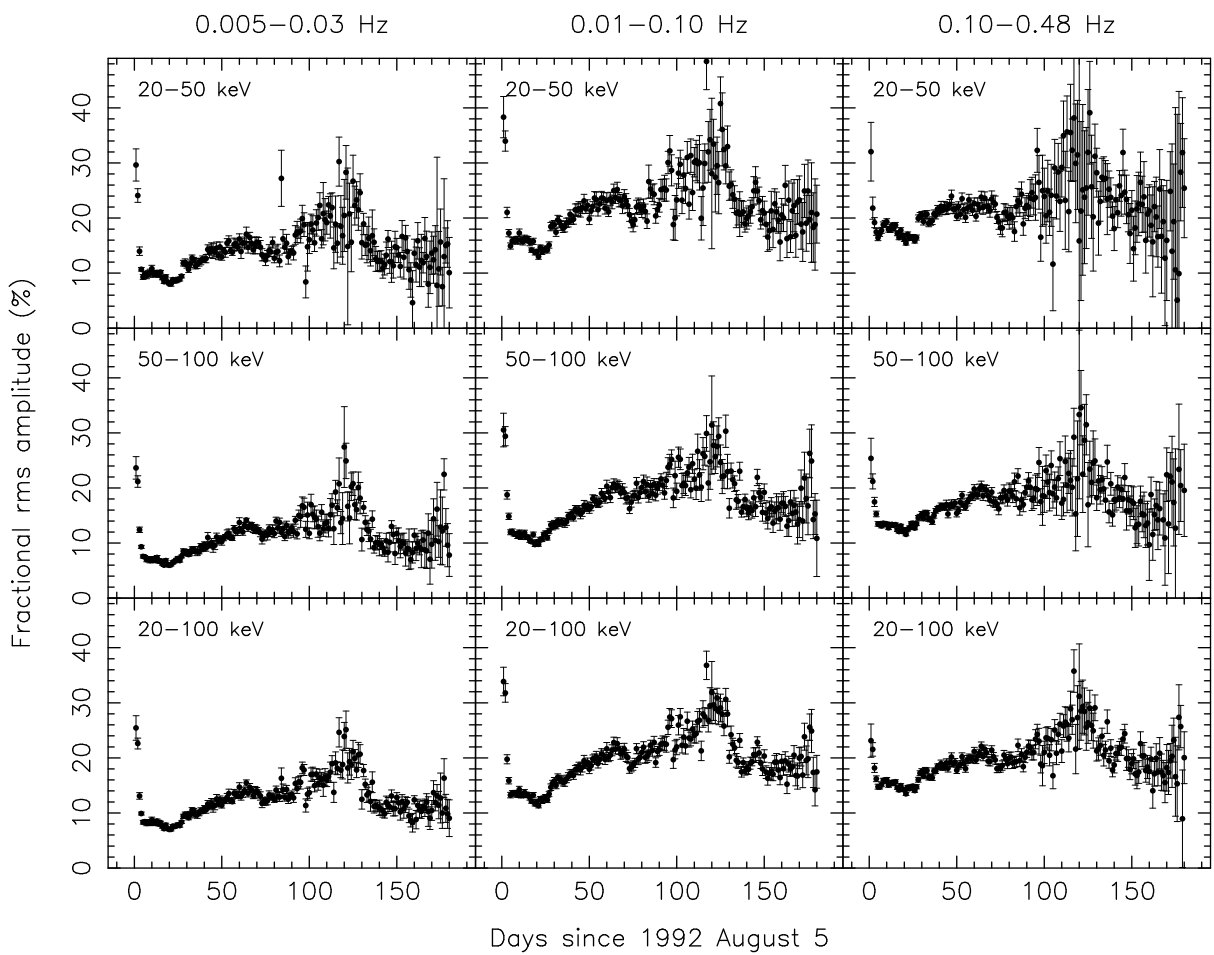,width=16cm,bbllx=47pt,bblly=80pt,bburx=553pt,bbury=722pt,angle=-90}} 
\caption[]{
History of fractional rms amplitudes of \gro04\ in the 20--50 {\sl (top row)}, 50--100 
{\sl (middle row)}, and 20--100 keV {\sl (bottom row)} energy band during 180 days following the 
start of the X-ray outburst on 1992 August 5. The single-day averaged PDSs were integrated in the 
0.005--0.03 {\sl (left column)}, 0.01--0.10 {\sl (middle column)}, and 0.10--0.48 Hz {\sl (right 
column)} frequency intervals. The errors were calculated from the sample variances.
} 
\label{0422_fig9}
\end{figure*}

We determined fractional rms amplitudes by integrating the single-day averaged PDSs of \gro04\ 
in the 20--50, 50--100, and 20--100 keV energy bands over three different frequency intervals. 
One interval 
covered the flat part of the power spectrum below the break frequency (0.005--0.03 Hz; 12 bins).  
A second interval was selected in the power-law tail of the PDSs (0.10--0.48 Hz; 199 bins), and the 
last frequency interval was chosen across the break frequency (0.01--0.10 Hz; 46 bins), including 
both the flat part and the power-law tail of the PDSs. 
The history of the integrated fractional rms amplitudes of \gro04\ in time are given in 
Figure~\ref{0422_fig9}. For each frequency interval over which the PDSs were integrated, the general 
shapes of these curves in the three different energy bands are very similar. The largest rapid X-ray 
variability occurs in the lowest energy band.

During the onset of the primary X-ray outburst of \gro04, the fractional rms amplitudes decrease 
rapidly over the total frequency interval in each of the three energy bands of the daily averaged 
PDSs. The fractional rms amplitudes are especially large during the first two days of the outburst 
in the low frequency intervals, which can also be seen in the dynamical power spectrum of  
Fig.~\ref{0422_fig3}. At the X-ray maximum of \gro04 and shortly thereafter, the rms amplitudes continue 
to decrease, but at a slower pace. About two weeks after X-ray maximum, the rms amplitudes 
reach an absolute minimum in each of the three energy bands and in each frequency interval. 
From that moment on, the fractional rms amplitudes gradually increase until day $\sim$\,65 of the outburst, 
after which they suddenly decrease to a local minimum at day $\sim$\,73. 
Such variations in the fractional rms amplitude, are present in each of the histories 
displayed in Fig.~\ref{0422_fig9}. 
The flux history of \gro04\ does not show any features during this phase of the outburst, 
but proceeds its smooth exponential decay (see Fig.~\ref{0422_fig1}). 

After the local minimum in fractional rms amplitude at day $\sim$\,73 of the outburst, the amplitudes 
increase again, but become uncertain and are dominated by detector noise at low flux levels 
of \gro04\ due to unresolved sources in the uncollimated field of view of BATSE. The largest fractional 
rms amplitudes are obtained shortly before the onset of the secondary maximum in the light curve. 
At this secondary maximum, the fractional rms amplitudes obtained in the 0.01--0.10 and 0.10--0.48 Hz 
frequency intervals again reach a local minimum. The fractional rms amplitudes obtained in the 0.001--0.03 
Hz interval do not exhibit such a minimum, but appear to be flat during the remainder of our analysis.

\begin{figure*} 
\centerline{\psfig{figure=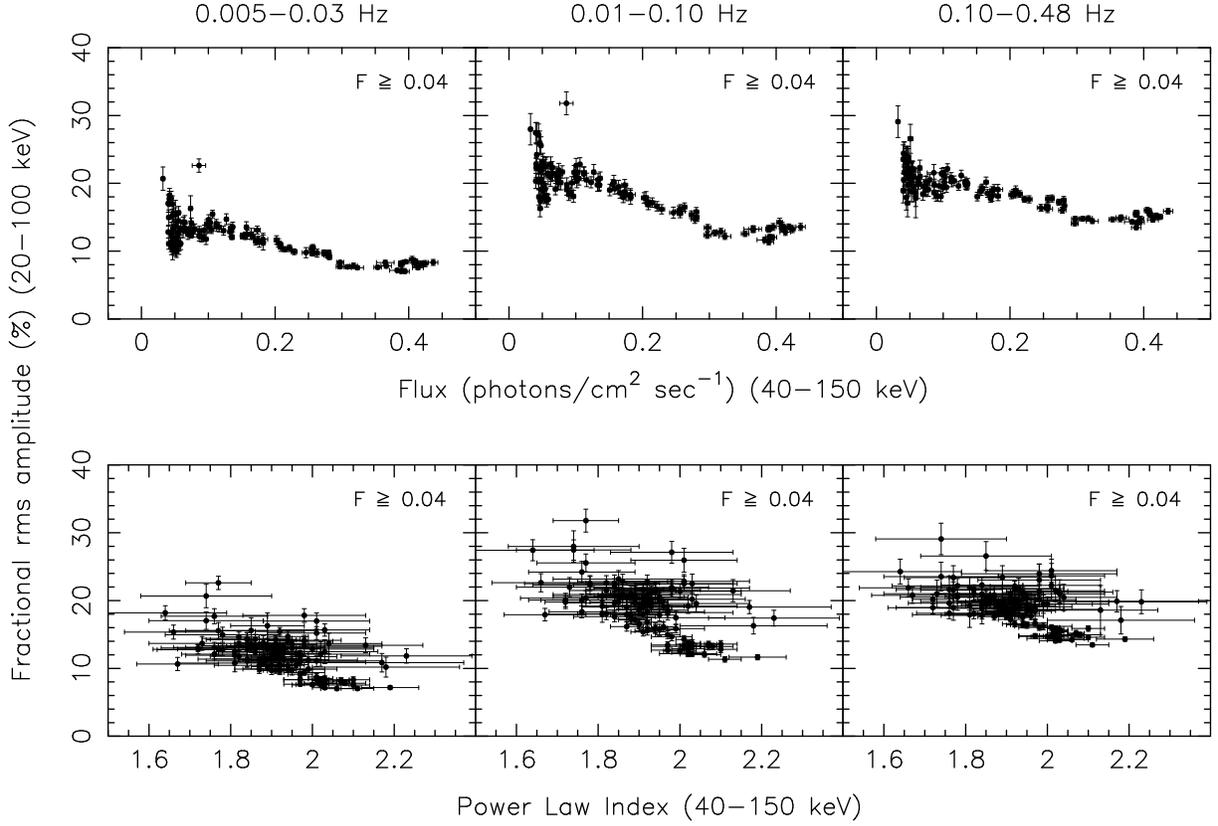,width=16cm,bbllx=47pt,bblly=80pt,bburx=486pt,bbury=722pt,angle=-90}} 
\caption[]{
Fractional rms amplitudes of \gro04\ in the 20--100 keV energy band integrated in the 0.005--0.03 
{\sl (left column)}, 0.01--0.10 {\sl (middle column)}, and 0.10--0.48 Hz {\sl (right column)} 
frequency intervals, versus the flux {\sl (top row)} and photon power-law index {\sl (bottom row)} 
in the 40--150 keV energy band. The sample is flux-limited 
({\it F}$\geq$0.04 photons/cm$^{\rm 2}$ sec$^{-1}$) and consists of 103 data points of the primary, and 23 
data points of the secondary outburst.
} 
\label{0422_fig10}
\end{figure*}

We have plotted the 20--100 keV fractional rms amplitudes obtained in the 0.005--0.03, 0.01--0.10 
and 0.10--0.48 Hz frequency intervals, versus the 40--150 keV flux and photon power-law index of 
\gro04\ in Figure~\ref{0422_fig10}. The distribution is flux-limited, similar to 
Fig.~\ref{0422_fig2} ($F\geq$0.04 photons/cm$^{\rm 2}$ sec$^{-1}$), and consists of 103 data points 
of the primary, and 23 data points of the secondary outburst. 
The fractional rms amplitudes appear to be anticorrelated with the flux of \gro04; the smallest rms 
amplitudes were obtained at the highest flux values. 
The single point deviating from the general trend in the fractional rms amplitude versus flux distribution 
in the 0.005--0.03 and 0.01--0.10 Hz frequency intervals corresponds to day 2 of the outburst. 
Again, this suggests that the low frequency X-ray variability of \gro04\ was exceptionally high during the start 
of its outburst. 
The photon power-law indices and fractional rms amplitudes also appear to be anticorrelated: the trend in 
Fig.~\ref{0422_fig10} is one of larger fractional rms amplitudes as the X-ray spectrum of \gro04\ hardens.

\section{Lag spectra}

We have calculated lags between the X-ray variations in the 20--50 and 50--100 keV energy 
bands of the 1.024 sec time resolution data of \gro04. The cross amplitudes were created 
from the Fourier amplitudes 
$a_j^l$\,$=$\,$\sum_k c_k^l \exp(i 2 \pi k j/n)$, where $n$ is the 
number of time bins (512), $c_k^l$ denotes the count rate in bin 
$k$\,$=$\,$0, \cdots, n-1$ and channel number $l$\,$=$\,0, 1 and $j$\,$=$\,$-n/2, \cdots, n/2$ 
corresponds to Fourier frequencies $2\pi j/n\Delta T$. The complex cross 
spectra of channel 0 and 1 are given by $C_{j}^{12}$\,$=$\,$a^{2\ast}_j a^1_j$ and were 
averaged daily. Errors on the real and imaginary parts of these daily averaged 
cross spectra were calculated from the respective sample variances, and 
formally propagated when computing the phase and time lags. The phase lags as a 
function of frequency are obtained from the cross spectra via 
$\phi_{j}$\,$=$\,arctan[Im($C_{j}^{12}$)/Re($C_{j}^{12}$)], and the time lag 
$\tau_{j}$\,$=$\,$\phi_{j}/2\pi\nu_{j}$, with $\nu_{j}$ the frequency in Hz of the 
$j$-th frequency bin. With these definitions, lags in the hard (50--100 keV) 
with respect to the soft (20--50 keV) X-ray variations, appear as positive 
angles. 

\begin{figure*} 
\centerline{\psfig{figure=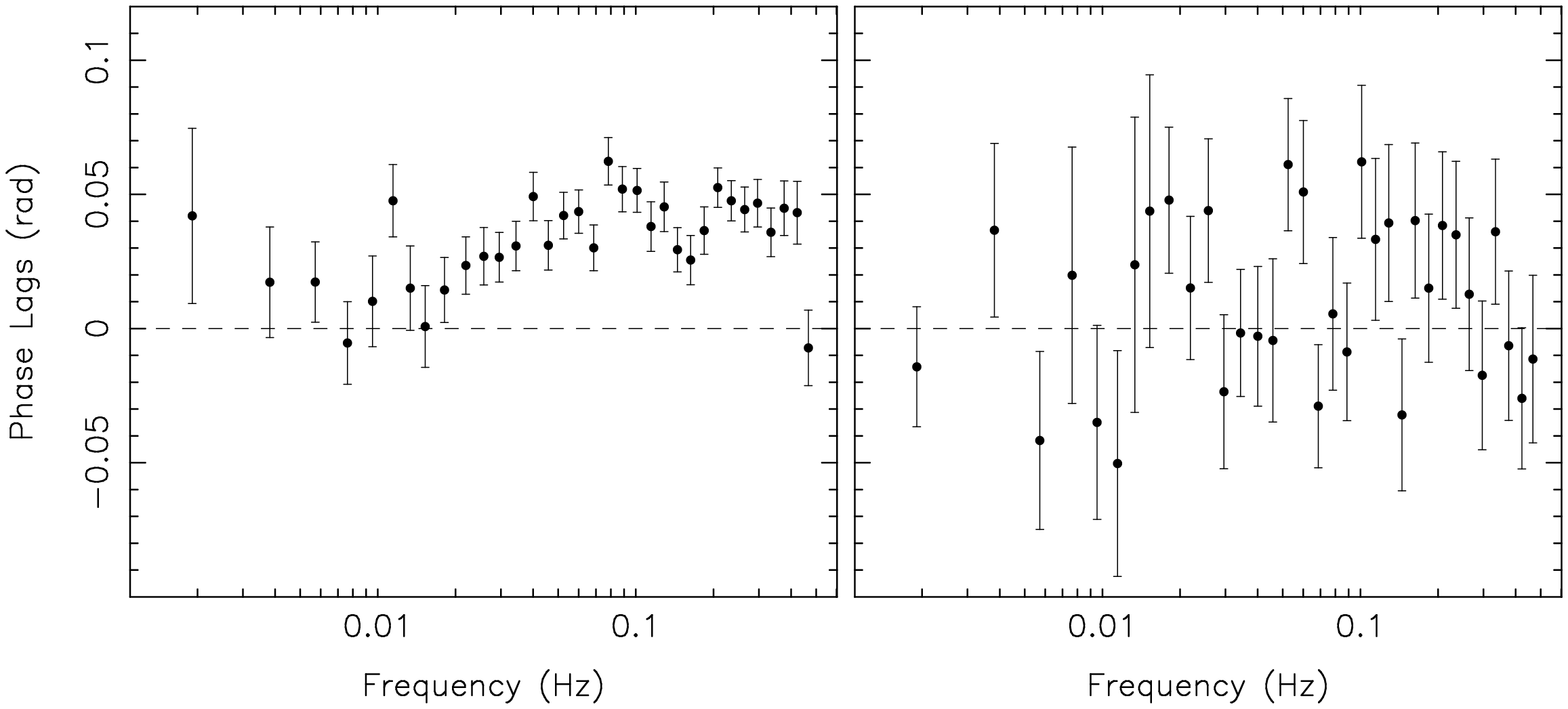,width=16cm,bbllx=238pt,bblly=94pt,bburx=537pt,bbury=758pt,angle=-90}} 
\caption[]{
Average phase lags of \gro04\ between the 20--50 and 50--100 keV energy bands (hard lags appear as positive 
angles). 
The left panel consists of a 30 day average at the start of the outburst, while the right panel 
displays a 95 day average of a flux-limited sample ({\it F}$\geq$0.04 photons/cm$^{\rm 2}$ sec$^{-1}$) of the 
remaining data. 
The frequency scale has been logarithmically rebinned into 34 bins.
} 
\label{0422_fig11} 
\end{figure*}

\begin{figure*} 
\centerline{\psfig{figure=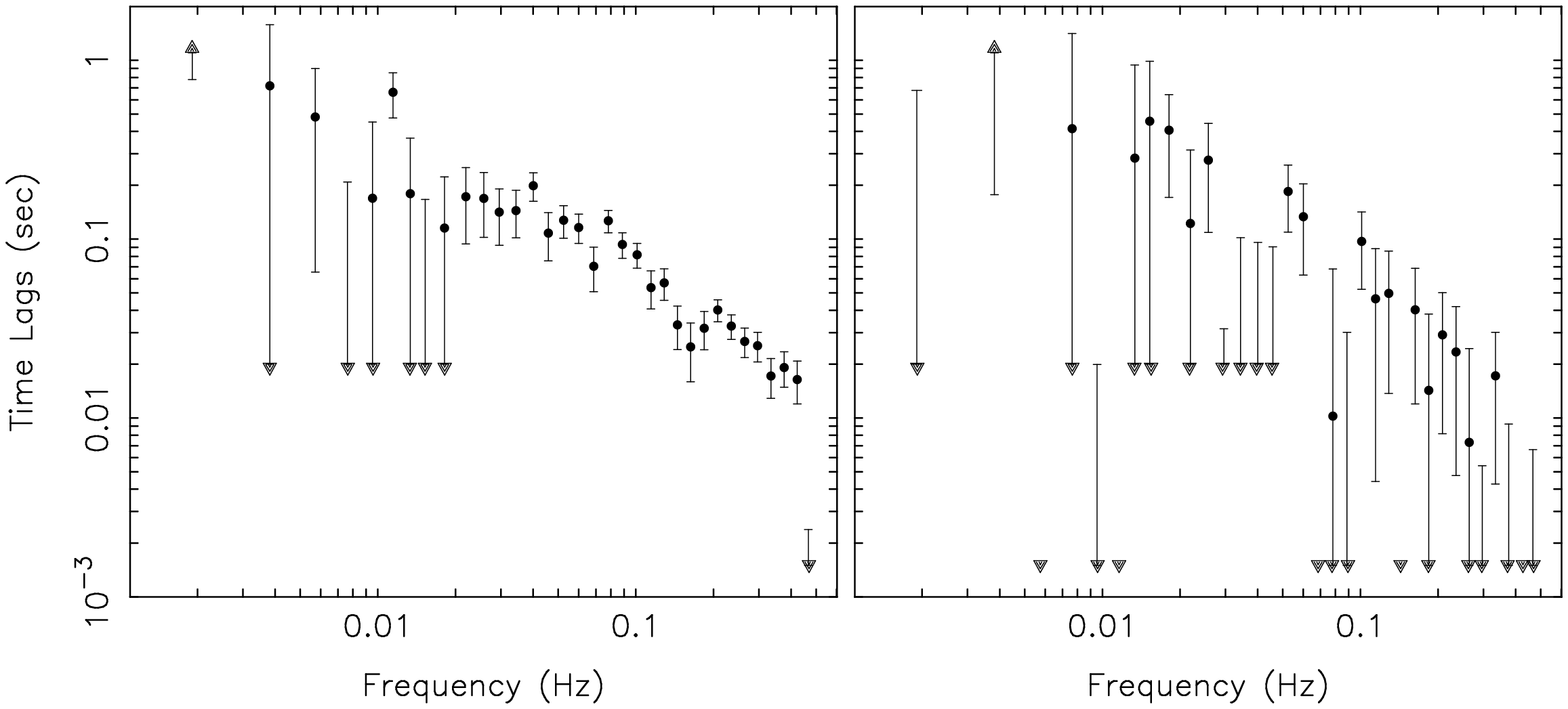,width=16cm,bbllx=238pt,bblly=94pt,bburx=537pt,bbury=758pt,angle=-90}} 
\caption[]{
Average time lags of \gro04\ between the 20--50 and 50--100 keV energy bands (hard lags appear as positive delays). 
The left panel consists of a 30 day average at the start of the outburst, while the right panel 
displays a 95 day average of a flux-limited sample ({\it F}$\geq$0.04 photons/cm$^{\rm 2}$ sec$^{-1}$) of the 
remaining data. The frequency scale has been logarithmically rebinned into 34 bins. 
Upper (lower) limits are indicated by triangles.
}
\label{0422_fig12} 
\end{figure*}

The phase lags of \gro04\ between the X-ray variations in the 20--50 and 50--100 keV 
energy bands, averaged over a 30 day interval at the start of the outburst, and averaged 
over an interval covering the following 95 days, are presented in Figure~\ref{0422_fig11}. 
Cross spectra for a large number of days must be averaged and converted to lag values to obtain 
sufficiently small errors. Cross spectra of the final 55 days of our analysis were not taken 
into account in the second average, as inclusion of these data did not significantly improve 
the quality of the average cross spectrum. 
Figure~\ref{0422_fig12} shows the corresponding time lags on a logarithmic scale. Lags at 
frequencies above 0.5$\nu_{\rm Nyq}$ are displayed but not taken into account in our analysis, 
as Crary \ea\ (1998) have shown that lags between 0.5$\nu_{\rm Nyq}$ and $\nu_{\rm Nyq}$ can be 
affected by data binning, and therefore, decrease artificially to zero. 
Our results show that at the lowest frequencies the phase lags are consistent with zero 
(0.014$\pm$0.006 rad, 0.001--0.02 Hz; 9 bins).  
At frequencies $\geq$0.02 Hz, the hard X-rays lag the soft by 0.039$\pm$0.003 rad 
(average of 0.02--0.20 Hz; 94 bins) during the 30 days following start of the outburst of \gro04. 
The phase lag derived for the following 95 days (0.017$\pm$0.007 rad, 0.02--0.20 Hz) is not statistical 
significant. 
During the first 30 days of the outburst of \gro04, the hard X-rays lag the soft by an amount of 0.02--0.2 
sec in the frequency interval 0.02--0.20 Hz.
The time lags decrease in this period with frequency as a power law, with index 0.88$\pm$0.04 for frequencies 
$\geq$\,0.01 Hz. 
Although primarily consisting of upper limits, the time lags in the 95 day average are consistent with 
those obtained at the beginning of the outburst (power-law index of 1.04$\pm$0.32 for $\nu\geq$0.01 Hz).

\section{Discussion}

Soon after the X-ray detection of \gro04, a possible optical counterpart was identified by 
Wagner \ea\ (1992) and Castro-Tirado \ea\ (1992, 1993). 
During the decay to quiescence, its optical brightness was reported to be modulated at several 
periods ranging between 2.1 and 16.2 hrs. 
Filippenko \ea\ (1995) determined the mass function and orbital period to be 
1.21$\pm$0.06 M$_{\odot}$, and 5.08$\pm$0.01 hrs, respectively. This relatively low value 
for the mass function was confirmed by Orosz \& Bailyn (1995) and Casares \ea\ (1995). 
The mass function of \gro04\ is one of the lowest measured values for the SXTs analyzed to 
date, and provides by itself no dynamical evidence that the compact star in \gro04\ is a black hole. 
Callanan \ea\ (1996) derived an upper limit to the orbital inclination of 45$^\circ$ from the ellipsoidal 
variations. 
Combined with the spectroscopic measurements, this limiting 
inclination implies a mass of the compact star in \gro04\ in excess of 3.4 M$_{\odot}$. 
The limits put to the inclination of \gro04\ by Beekman \ea\ (1997) from infrared and optical photometry 
(10$^\circ$--31$^\circ$) imply an even higher mass limit of the compact star: $\ga$\,9 M$_{\odot}$.
Therefore, based on dynamical arguments, it can be concluded that \gro04\ most probably contains 
a black hole. 
Thus, the dynamical evidence on \gro04\ supports the early suggestion (Roques \ea\ 1994), made on the 
basis of the X-ray properties, that this system contains a black hole. 

It is not possible on the basis of BATSE observations alone, to distinguish between black hole source 
states. 
However, two independent observations at low X-ray energies during the decay of the X-ray light curve 
of \gro04, suggest that it was in the low (or hard) state. 
The X-ray spectrum of \gro04, obtained about 24 days after the start of the outburst by TTM (2--30 keV) 
and HEXE (20--200 keV) on board Mir-Kvant, had a power-law shape (photon index 1.5), with no strong 
soft component and an exponential cutoff at energies above 100 keV (Sunyaev \ea\ 1993). 
ROSAT HRI (0.1--2.4 keV) observations $\sim$\,42 days after the start of the outburst, also show no 
indication for an ultrasoft excess in the X-ray spectrum (Pietsch \ea\ 1993).  
Therefore, these observations show that 24--42 days after the X-ray outburst had started, \gro04\ was 
in the low state. The lack of significant changes in the hard X-ray properties 
(see Section~\ref{lc_spectrum}), indicate that this conclusion applies to the whole outburst.

\subsection{Comparison to Cyg X-1}

From an analysis of approximately 1100 days of BATSE data (20--100 keV) of the BHC Cyg X-1, 
covering the period 1991--1995, Crary \ea\ (1996a) 
found a strong correlation between the spectral slope and both the high energy X-ray flux and 
the variability thereof. Although low-energy coverage was lacking, it is likely that Cyg X-1 was 
in the low state during almost the entire period based on its strong rapid X-ray variability and the 
presence of a hard spectral component (Crary \ea\ 1996b). 
A possible transition to the high state may have lasted for $\sim$\,180 days, starting 1993 
September, as the source flux gradually declined over a period of 150 days to a very low level. 
After that, the flux rose within 30 days, back to the level it approximately had before the low 
flux episode occured (Crary \ea\ 1996b). 

\begin{figure*} 
\hbox{
\psfig{figure=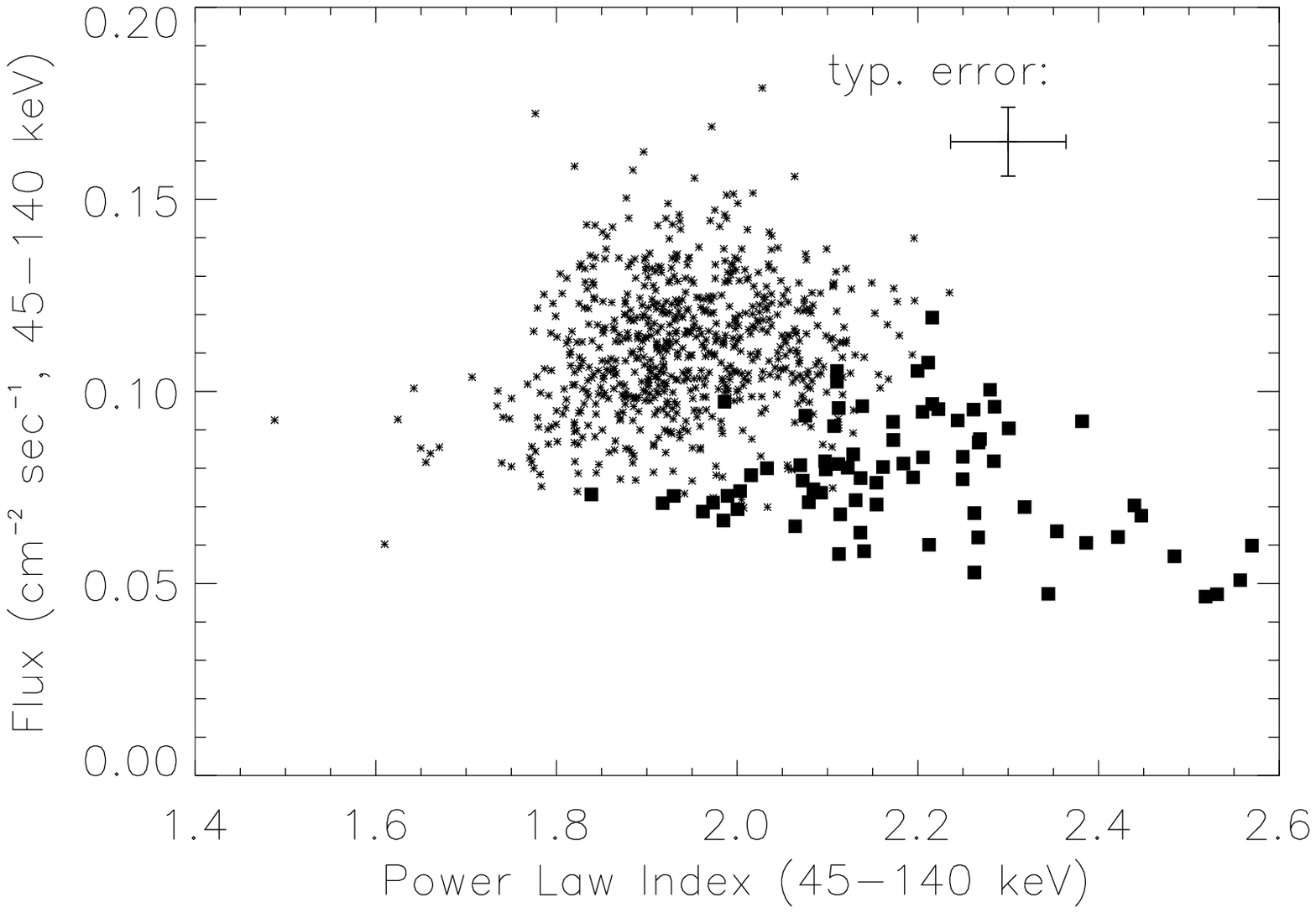,width=7.7cm,bbllx=82pt,bblly=373pt,bburx=542pt,bbury=692pt}
\hspace{8.2cm}
}
\vspace{5mm}
\hbox{
\psfig{figure=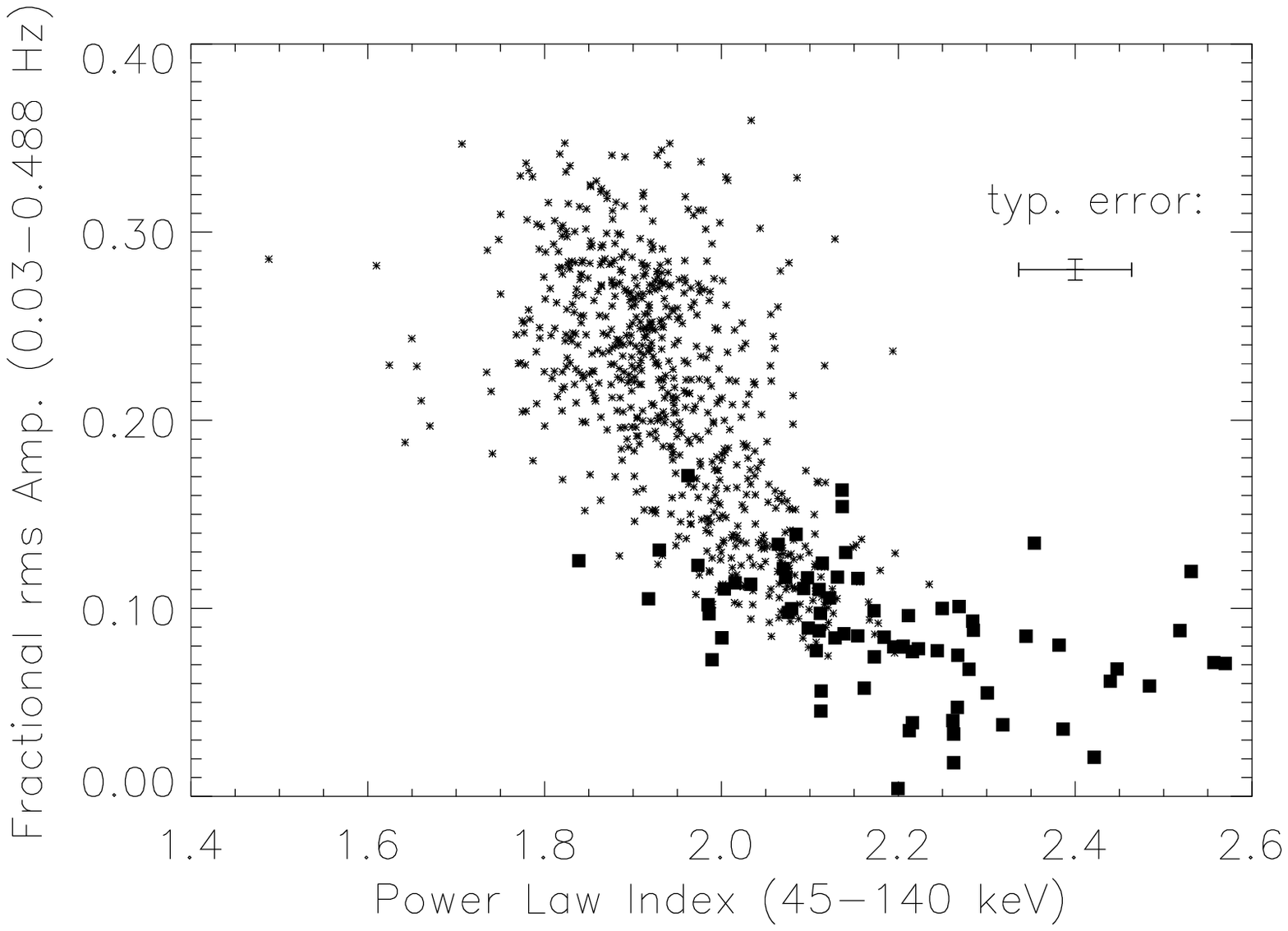,width=7.7cm,bbllx=83pt,bblly=374pt,bburx=546pt,bbury=705pt} 
\hspace{0.4cm}
\psfig{figure=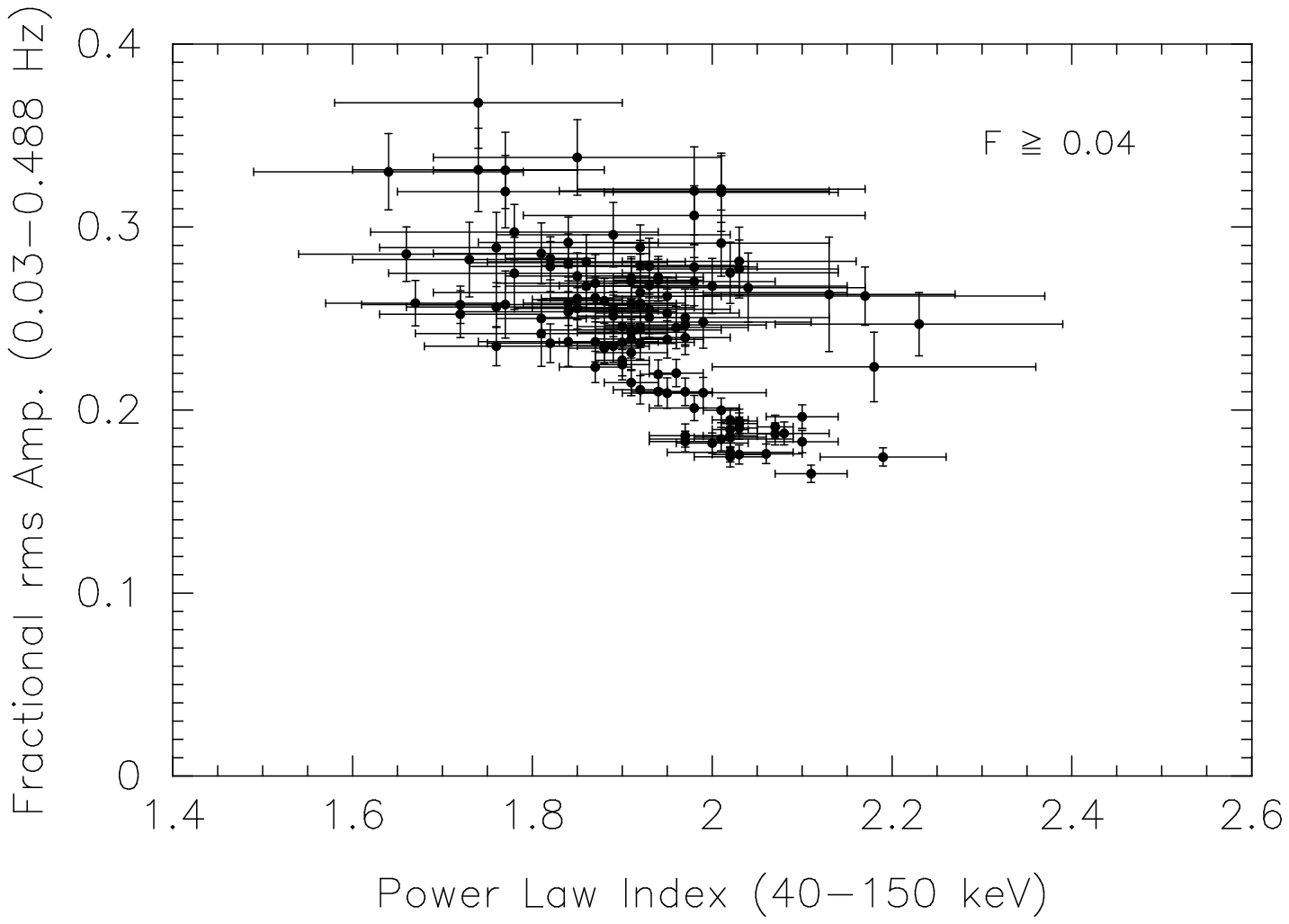,width=7.7cm,bbllx=217pt,bblly=81pt,bburx=509pt,bbury=485pt,angle=-90}
}
\vspace{5mm}
\hbox{
\psfig{figure=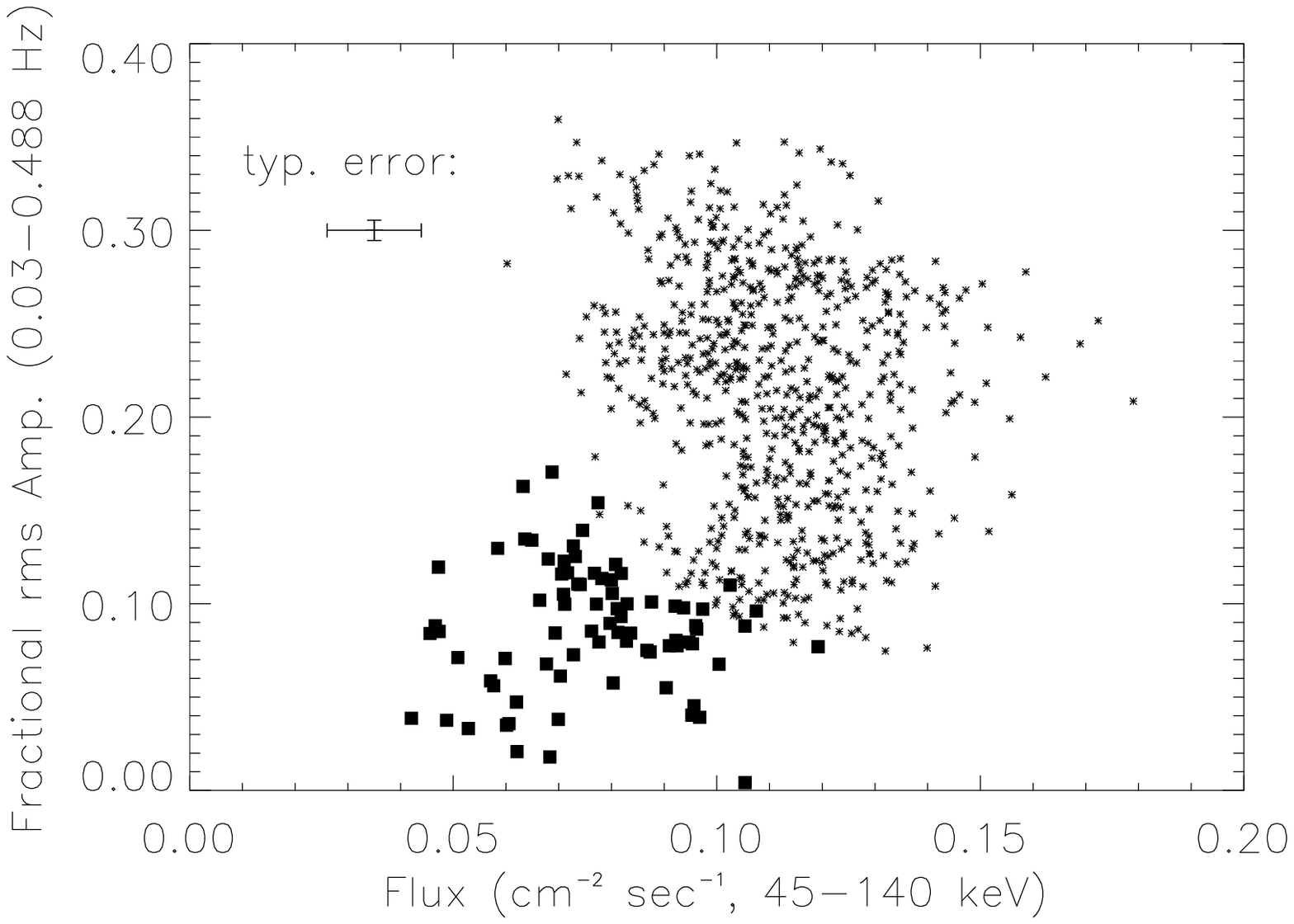,width=7.7cm,bbllx=83pt,bblly=374pt,bburx=540pt,bbury=705pt}
\hspace{0.4cm}
\psfig{figure=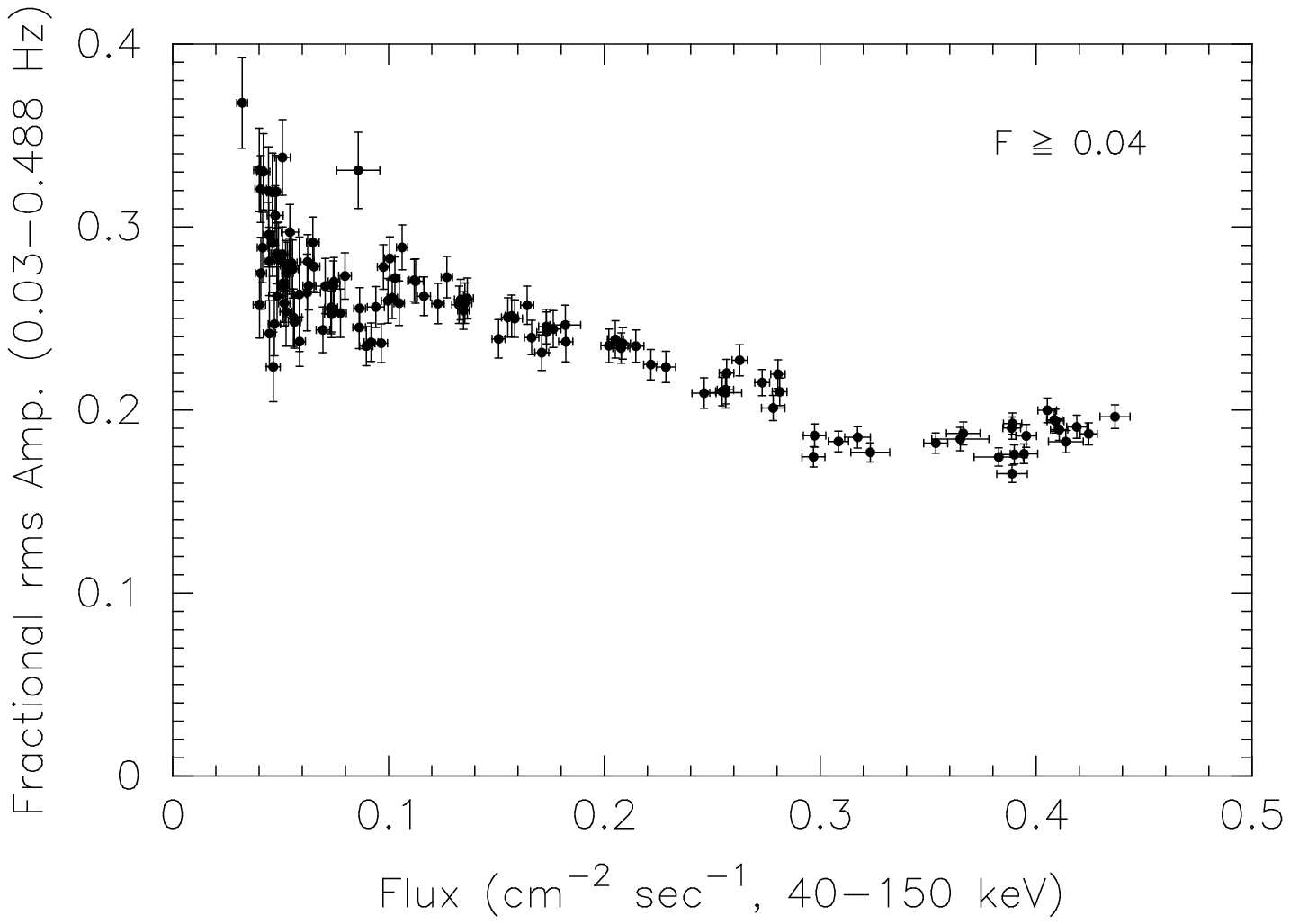,width=7.7cm,bbllx=217pt,bblly=81pt,bburx=509pt,bbury=485pt,angle=-90}
}
\caption[]{
Cyg~X-1 45--140 keV flux (cm$^{-2}$ sec$^{-1}$) versus photon power-law index from a fit in the 
45--140 keV range {\sl (top-left)}. 
Fractional rms amplitude (0.03--0.488 Hz) in the 20--100 keV band versus photon power-law index 
of Cyg~X-1 {\sl (middle-left)}, and \gro04\ {\sl (middle-right)}. 
Fractional rms amplitude (0.03--0.488 Hz) in the 20--100 keV band versus flux of 
Cyg~X-1 {\sl (bottom-left)} and \gro04\ {\sl (bottom-right)}. 
Figures of Cyg~X-1 are taken from Crary \ea\ 1996a.
}
\label{0422_fig13} 
\end{figure*}

The strong correlation of the flux and power-law index of \gro04\ in the 40--150 keV energy band, 
shown in Fig.~\ref{0422_fig2}, is different from the correlation between the same quantities in Cyg X-1 as 
observed by Crary \ea\ (1996a). 
For flux values above 0.20 photons/cm$^{\rm 2}$ sec$^{-1}$, the flux and photon power-law index of 
\gro04\ are related linearly; the larger the high-energy X-ray flux, the softer the X-ray spectrum. 
The power-law index is constant at 1.89$\pm$0.02 for {\it F}$<$0.20 photons/cm$^{\rm 2}$ sec$^{-1}$. 
The correlation between the flux and power-law index of Cyg X-1 in the 45--140 kev energy band, however, 
is in the opposite direction; the larger the 45--140 keV flux of Cyg X-1, the harder its X-ray spectrum 
(Crary \ea\ 1996a).  
For comparison, the figures of Crary \ea\ (1996a) are reproduced in Figure~\ref{0422_fig13}. These show the 
correlations between the fractional rms amplitude (0.03--0.488 Hz) in the 20--100 keV energy band, and the 
photon power-law index and flux, for both \gro04\ and Cyg~X-1.

The correlations between the photon power-law index and fractional rms variability in Cyg X-1 and 
\gro04\ are very similar; for both sources, the lower the photon power-law index, the larger the fractional 
rms amplitude. 

As a result, the correlation between the flux and the fractional rms amplitude in the 20--100 keV energy band 
are entirely different for Cyg X-1 and \gro04. In the Cyg X-1 data, the distribution of flux and 
fractional rms amplitude measurements (frequency interval 0.03--0.488 Hz) follows a broad, upturning 
band, i.e., larger fractional amplitudes at higher flux values (Crary \ea\ 1996a). 
However, a strong anticorrelation is found between the 20--100 keV fractional rms amplitude (both 
0.01--0.10 and 0.10--0.48 Hz) of \gro04, and its 40--150 keV flux (see, e.g., Fig.\ref{0422_fig10}). 
This distribution traces a narrow path in the diagram, with the smallest fractional amplitudes occurring at 
the largest flux values, or equivalently, at the start of the outburst of \gro04.

Crary \ea\ (1998) studied the Fourier cross spectra of Cyg X-1 with BATSE during a period of 
almost 2000 days. During this period, Cyg X-1 was likely in both the low, and high or intermediate state. 
Crary \ea\ (1998) found that the lag spectra between the X-ray variations in the 20--50 and 50--100 keV 
energy bands of Cyg X-1 do not show an obvious trend with source state.
The X-ray variations of Cyg X-1 in the 50--100 keV energy band lag those in the 20--50 keV energy band 
over the 0.01--0.20 Hz frequency interval by a time interval proportional to $\nu^{-0.8}$ (Crary \ea\ 1998). 
The general shape and sign of the phase and time lag spectra of \gro04\ are very similar to those of Cyg X-1. 
Significant phase (or equivalently, time) lags in the \gro04\ data could only be derived during the 
early stage of its outburst. During this period, the variations in the 50--100 keV energy band lag 
those in the 20--50 keV energy band by 0.039(3) rad in the frequency interval 0.02--0.20 Hz. 
The time lags of \gro04\ during the first 30 days of its outburst, decrease with frequency as a power law, 
with index $\sim$\,0.9 for $\nu>$0.01 Hz.

\subsection{Break frequency}

In several BHCs the break frequency of the PDSs has been observed to vary by up to an order of magnitude
while the high frequency part remained approximately constant (Belloni \& Hasinger 1990a; Miyamoto \ea\ 1992). 
As a result, the break frequency and power density at the break are strongly anticorrelated. 
M\'{e}ndez \& van der Klis (1997) 
have collected the relevant data on all BHC power spectra in the low, intermediate and very high state 
for six BHCs, and show that among these sources the break frequency is anticorrelated 
to the power density at the break. 
As this correlation appears to hold across different source states, M\'{e}ndez \& van der Klis (1997) 
suggest a correlation with mass accretion rate may exist, i.e. the break frequency increases (and the 
power density decreases) with increasing mass accretion rate (see also van der Klis 1994a). 

The break frequency and power density at the break in the PDSs of \gro04\ are clearly not anticorrelated 
(see Fig.~\ref{0422_fig7}). 
The power density at the break in the 20--100 keV energy band ranged between 44 and 89\% Hz$^{-1/2}$, while the 
break frequency was rather constant, at either 0.041$\pm$0.006 or 0.081$\pm$0.015 Hz. 
All PDSs in which the break of the power spectrum was detected near $\sim$\,0.04 Hz were 
obtained during the primary X-ray outburst of \gro04; 
all PDSs obtained during the secondary outburst have a break frequency near $\sim$\,0.08 Hz. 
Therefore, the relation between the break frequency and power density at the break is different 
from the one observed in Cyg X-1 (Belloni \& Hasinger 1990a; Miyamoto \ea\ 1992). 
This contrast between the PDSs of \gro04\ and Cyg X-1 was also reported by Grove \ea\ (1994) based on 
OSSE data of the first stage of the X-ray outburst of \gro04. 

\begin{figure*} 
\centerline{\psfig{figure=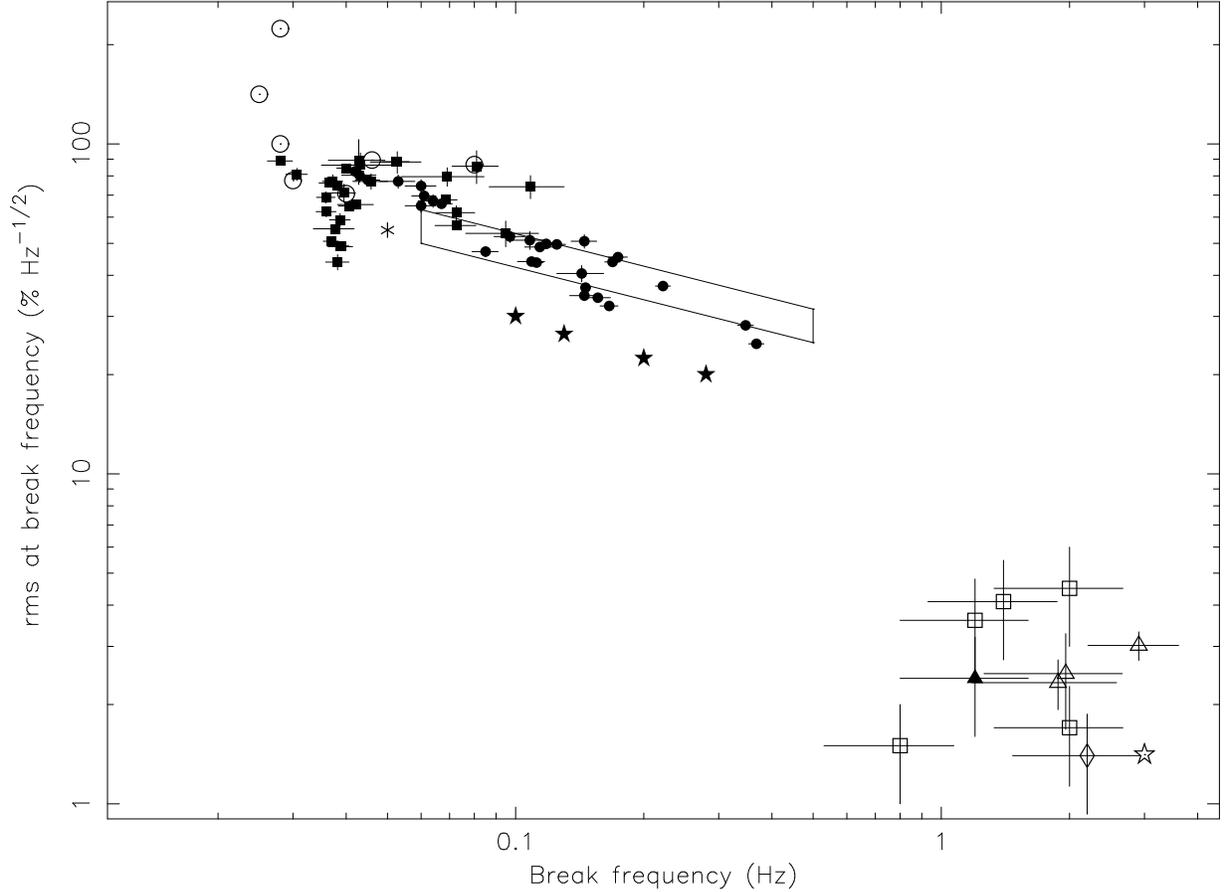,width=16cm,bbllx=78pt,bblly=43pt,bburx=569pt,bbury=700pt,angle=-90}} 
\caption[]{
Relation between break frequency and power density at the break, taken from M\'{e}ndez \& van der Klis 
(1997), with the addition of the \gro04\ data. 
{\sl Filled circles:} Cyg X-1 in the LS (Belloni \& Hasinger (1990a); 
{\sl filled stars:} GRO~J1719$-$24 in the LS (van der Hooft \ea\ 1996, 1998); 
{\sl open circles:} GS~2023$+$338 in the LS (Oosterbroek \ea\ 1997); 
{\sl asterisk:} \gro04\ in the LS (Grove \ea\ 1994); 
{\sl filled squares:} \gro04\ in the LS (this paper); 
{\sl open squares:} GX~339$-$4 in the VHS (Miyamoto \ea\ 1991, 1993); 
{\sl open diamond:} GS~1124$-$68 in the VHS (Miyamoto \ea\ 1993); 
{\sl filled triangle:} GX~339$-$4 (Belloni \& Hasinger 1990b); 
{\sl open star:} GS~1124$-$68 in the IS (Belloni \ea\ 1997); 
{\sl open triangles:} GX~339$-$4 in the IS (M\'{e}ndez \& van der Klis 1997). 
The marked region corresponds to Cyg~X-1 LS data from Crary \ea\ (1996a). 
}
\label{0422_fig14} 
\end{figure*}

The break frequency and power density at the break of BHC power spectra in the low, intermediate and very high 
state (LS, IS, VHS) collected by M\'{e}ndez \& van der Klis (1997) are reproduced in Figure~\ref{0422_fig14}, 
together with the 20--100 keV data of \gro04. 
However, this representation should be taken with caution, as it combines data obtained in different energy bands, 
and data of different source states. 
Fig.~\ref{0422_fig14} shows a clear anticorrelation of the break frequency and power density at the break over 
two orders of magnitude in both frequency and power density. 
The low state BATSE data (20--100 keV) of GRO~J1719$-$24 (van der Hooft \ea\ 1996) and Cyg~X-1 (Crary \ea\ 1996a), 
and low state EXOSAT data of Cyg~X-1 (Belloni \& Hasinger 1990a), clearly follows this anticorrelation. 
However, the power density at the break detected in the PDSs of GS~2023$+$338 is approximately constant for 
break frequencies in the 0.03--0.08 Hz interval, but shows a sudden turnover at frequencies $<$0.03 Hz to 
large power densities. 
The absence of the anticorrelation of the break frequency and the power density at the break in the power 
spectra of GS~2023$+$338 was first reported by Oosterbroek \ea\ (1997). 
Although these studies of \gro04\ and GS~2023$+$338 show that the relation between the break 
frequency and power density at the break differs in detail from other BHCs in the LS, the data of \gro04\ and 
GS~2023$+$338 displayed in Fig.~\ref{0422_fig14} does fit the general anticorrelation, observed in several sources 
over two decades of frequency and power density, within a factor of $\sim$\,2. 
It is interesting to note that the absence of an anticorrelation of the break frequency and power density at 
the break, is found in the PDSs of the two sources which show the lowest break frequencies of all BHCs: 
\gro04\ and GS~2023$+$338.

\subsection{Earlier temporal analyses of \gro04} 

The hard X-ray variability of \gro04\ has been studied by Grove \ea\ (1994) with OSSE (35--600 keV), and by 
Denis \ea\ (1994) and Vikhlinin \ea\ (1995) with SIGMA in the 40--300 and 40--150 keV energy band. 
The OSSE observations were obtained from 1992 August 11 until September 17, i.e., days 7--44 of the X-ray 
outburst. Grove \ea\ (1994) computed PDS in the 35--60 and 75--175 keV energy band. 
The power spectra show breaks at frequencies of $\sim$\,10$^{-2}$ Hz and a few Hz, and a strong peaked noise 
component at 0.23 Hz. Statistically significant noise is detected at frequencies above 20 Hz. 
The total fractional rms in the 0.01--60 Hz frequency interval for the entire OSSE pointing is 
$\sim$\,40\% (35--60 keV) and $\sim$\,30\% (75--175 keV), respectively. 

SIGMA observed \gro04\ from 1992 August 15 until September 25 (days 11--52). 
Vikhlinin \ea\ (1995) computed power spectra in the frequency interval 10$^{-4}$--10 Hz. 
The power density is nearly constant below a break frequency at 0.03 Hz, and decreases as a power law above, 
with index $\sim$\,0.9. A strong QPOs peak is detected at a frequency of 0.3 Hz. 
At the start of the SIGMA observation of \gro04\ (August 15--27), the PDSs show a significant decrease of the 
power densities for frequencies lower than 3.5$\times$10$^{-3}$ Hz. 
Such turnover of the PDSs at low frequencies is unusual; 
PDSs typically exhibit strong very-low frequency noise, or are constant for such low frequencies. 
The total fractional rms variations (40--150 keV) of \gro04\ in the frequency interval 10$^{-3}$--10$^{-1}$ Hz 
increased from $\sim$\,15\% at the start of the SIGMA observations to $\sim$\,25\% at the end. 
Denis \ea\ (1994) reported an increase of the fractional rms variation over the same period from $\leq$\,25\% 
(2$\sigma$ upper limit) to $\sim$\,50\% (150--300 keV) in the 
2$\times$10$^{-4}$--1.25$\times$10$^{-1}$ Hz frequency interval.

These results of Vikhlinin \ea\ (1995) and Grove \ea\ (1994) are consistent with the results presented here. 
The shape of the power spectrum of \gro04\ in the 0.01--0.5 Hz interval is essentially identical in the three 
studies, although different, but partly overlapping, energy bands have been used. 
The values of the break frequency and power-law index (both, below and above the break frequency) determined in 
the OSSE and SIGMA data, are consistent with those presented here. 
However, OSSE and SIGMA only covered the early phases of the X-ray outburst of \gro04, with the exception of its 
very start. Therefore, the significant change of the break frequency during later stages of the X-ray outburst 
was missed. 
Fractional rms amplitudes were determined on a daily basis in the OSSE data (0.01--60 Hz) and SIGMA data 
(3.6$\times$10$^{-4}$--8$\times$10$^{-2}$ Hz). 
At the beginning of the OSSE and SIGMA observations, the fractional rms amplitudes remained constant, 
or perhaps slightly decreasing, followed by a turnover and gradual increase, similar to the fractional rms 
amplitude histories displayed in Fig.~\ref{0422_fig9}. 
However, due to data gaps, the moment of turnover is difficult to determine in the OSSE and SIGMA data. 

Low-frequency QPOs (0.04--0.8~Hz) have been observed in the PDSs of several BHCs: Cyg~X-1, LMC~X-1, GX~339$-$4 
and GRO~J1719$-$24 (van der Klis 1995b; van der Hooft \ea\ 1996). 
These QPOs were observed while the sources were likely in the low state, with the exception of LMC~X-1 where a 0.08~Hz 
QPO was found while an ultrasoft component dominated the energy spectrum, showing that the source was in the high state. 
During our observations \gro04\ was probably in the low state and its PDSs showed a strong QPOs peak with a centroid 
frequency between 0.13 and 0.26 Hz (20--100 keV). 
Therefore, \gro04\ is the fourth BHC with shows low-frequency QPOs in its PDSs while in the low state.

Peaked noise components and QPOs peaks in the power spectra of \gro04\ have been reported by various groups. 
Kouveliotou \ea\ (1992, 1993) reported QPO peaks centered at $\sim$\,0.03 and 0.2 Hz in different BATSE 
energy bands covering 20--300 keV. 
Grove \ea\ (1994) detected a strong peaked noise component in the OSSE data (35--60, 75--175 keV) at a centroid 
frequency of 0.23 Hz (FWHM $\sim$\,0.2 Hz), and evidence for additional peaked components near 0.04 and 0.1 Hz 
with a day to day variable intensity. 
Vikhlinin \ea\ (1995) report a strong QPO peak in the SIGMA data (40--150 keV) at 0.31 Hz (FWHM 0.16 Hz), 
with a fractional rms variability of $\sim$\,12\%. 
Likely, the reports of strong noise components at a few 10$^{-1}$ Hz all refer to the same feature in the PDSs 
of \gro04; the strong QPO peak with a centroid frequency ranging between 0.13 and 0.26 Hz (20--100 keV), present 
in the power spectra of \gro04\ during the first $\sim$\,70 days of its outburst. 
The peaked noise at a few 10$^{-2}$ Hz reported by Kouveliotou \ea\ (1992, 1993) and Grove \ea\ (1994), is 
detected near the break frequency of the power spectra. 
These detections may be supported by Vikhlinin \ea\ (1992), 
who reported peaked noise at 0.035 Hz (FWHM 0.02 Hz) in 40--70 keV SIGMA data, obtained early in the SIGMA 
pointing. 
However, Kouveliotou \ea\ (1993) report that this peaked noise structure is only detected occasionally in the 
PDSs of \gro04. 
In the analysis presented here, we do not find significant evidence for such peaked noise 
structures at a few 10$^{-2}$ Hz, possibly due to our daily averaging routine, or the fact that these noise 
components occur close to the break frequency. 

\subsection{Comptonization models}

The power law hard X-ray spectral component of accreting BHCs, which dominate the X-ray emission in the 
low state, can be described well by Compton upscattering of low-energy photons by a hot electron gas 
(Sunyaev \& Tr\"{u}mper 1979). 
In such a case, the energy of the escaping photons on average increases with the number of scatterings, and therefore, 
with the time they reside in the cloud. 
Therefore, high-energy photons lag those with lower energies by an amount proportional to the photon travel time. 
If the hard X-rays are emitted from a compact region in the immediate vicinity of the black hole, the resulting 
time lags should be independent of Fourier frequency and of the order of milliseconds.

The hard X-ray photons (50--100 keV) of \gro04\ lag the low-energy photons (20--50 keV) by as much as $\sim$\,0.1--1 
sec at low frequencies. 
The time lags are strongly dependent on the Fourier frequency, and decrease roughly as $\nu^{-1}$.
Similar lag behaviour have been observed in Cyg~X-1 (Cui \ea\ 1997; Crary \ea\ 1998) and GRO~J1719$-$24 
(van der Hooft \ea\ 1998) while in the low state. 
Recently, Kazanas, Hua \& Titarchuk (1997) argued that the Comptonization process takes place in an extended 
non-uniform cloud around the central source. 
Such a model can account for the form of the observed PDS and energy spectra of compact sources. 
Hua, Kazanas \& Titarchuk (1997) showed that the time lags of the X-ray variability depend on the density 
profile of such an extended but non-uniform scattering atmosphere. 
Their model produces time lags between the hard and soft bands of the X-ray spectrum that increase with 
Fourier period, in agreement with the observations. 
Therefore, analysis of the hard time lags in the X-ray variability of black-hole candidates could provide information 
on the density structure of the accretion gas (Hua \ea\ 1997). 
The time lags observed in GRO~J0422$+$32, Cyg~X-1 and GRO~J1719$-$24 are quite similar and support the idea 
that the Comptonizing regions around these black holes are similar in density distribution and size.

However, the observed lags require that the scattering medium has a size of order 10$^3$ to 10$^4$ Schwarzschild radii. 
It is unclear how a substantial fraction of the X-ray 
luminosity, which must originate from the conversion of gravitational potential energy into heat close to the 
black hole, can reside in a hot electron gas at such large distances. 
This is a generic problem for Comptonization models of the hard X-ray time lags. 
Perhaps very detailed high signal-to-noise cross-spectral studies of the rapid X-ray variability of accreting BHCs, and 
combined spectro-temporal modeling can solve this problem.

\section{Conclusions} 

We have analyzed the hard X-ray variability of \gro04. 
The canonical anticorrelation between the break frequency and the power at the break observed 
in Cyg X-1 and other BHCs in the low state, is not present in the PDSs of \gro04. 
The relation between the photon power-law index of the X-ray spectrum and the amplitude of the X-ray 
variations of \gro04\ has similarities to that of Cyg~X-1; however, the relation between the hard 
X-ray flux and the amplitude of its variation is opposite to what has been found in Cyg~X-1.

\acknowledgments
FvdH acknowledges support by the Netherlands Foundation for Research in Astronomy with 
financial aid from the Netherlands Organisation for Scientific Research (NWO) under 
contract number 782-376-011. 
FvdH also thanks the ``Leids Kerkhoven--Bosscha Fonds'' for a travel grant. 
CK acknowledges support from NASA grant NAG-2560. 
JvP acknowledges support from NASA grants NAG5-2755 and NAG5-3674. 
MvdK gratefully acknowledges the Visiting Miller Professor Program 
of the Miller Institute for Basic Research in Science (UCB). 
This project was supported in part by NWO under grant PGS 78-277.

%================= References ===============================

\clearpage
%
%%=============================================================
%       References
%%=============================================================

% 
%%============================================================= 
% 	Figures
%%============================================================= 

\end{document}